\documentclass[fleqn,usenatbib]{mnras}
\usepackage{newtxtext,newtxmath}
\usepackage[T1]{fontenc}
\usepackage{ae,aecompl}

\usepackage{subfig,graphicx,caption}
\usepackage{amsmath}	    % Advanced maths commands

\graphicspath{{./plots/}}

\title[GL of GWs: ML probability in galaxy lenses]
{Gravitational Lensing of Gravitational Waves: Probability of Microlensing in 
Galaxy-Scale Lens Population}

\author[A. K. Meena et al]{
Ashish Kumar Meena$^{1}$\thanks{E-mail: ashishmeena766@gmail.com},
Anuj Mishra$^{2}$,
Anupreeta More$^{2,3}$,
Sukanta Bose$^{2,4}$
and Jasjeet Singh Bagla$^{5}$
\\
\\
$^{1}$Physics Department, Ben-Gurion University of the Negev, P.O. Box 653,
Be'er-Sheva 84105, Israel 
\\
$^{2}$The Inter-University Centre for Astronomy and Astrophysics, 
Post Bag 4, Ganeshkhind, Pune 411007, India
\\
$^{3}$Kavli Institute for the Physics and Mathematics of the Universe, 
5-1-5 Kashiwanoha, Kashiwa-shi, Chiba 277-8583, Japan
\\
$^{4}$Department of Physics \& Astronomy, Washington State University, 
Pullman, WA 99164, USA
\\
$^{5}$Department of Physical Sciences, Indian Institute of Science 
Education and Research Mohali, Knowledge City, Sector 81, SAS Nagar, 
Punjab 140306, India
}
\date{Accepted XXX; Received YYY; in original form ZZZ}

\pubyear{2022}

\begin{document}
\label{firstpage}
\pagerange{\pageref{firstpage}--\pageref{lastpage}}
\maketitle

%%%%%%%%%%%%%%%%%%%%%%%%%%%%%%%%%%%%%%%%%%%%%%%%%%%%%%%%%%%%
\begin{abstract}
With the increase in the number of observed gravitational wave (GW) 
signals, detecting strongly lensed GWs by galaxies has become a real 
possibility. 
Lens galaxies also contain microlenses (e.g., stars and black holes), 
introducing further frequency-dependent modulations in the strongly 
lensed GW signal within the LIGO frequency range. 
The multiple lensed signals in a given lens system have different 
underlying macro-magnifications~($|\mu|$) and are located in varied 
microlens densities~($\Sigma_\bullet$), leading to different levels 
of microlensing distortions. 
This work quantifies the fraction of strong lens systems affected by 
microlensing using realistic mock observations. 
We study 50 quadruply imaged systems (quads) by generating 50 
realizations for each lensed signal. 
However, our conclusions are equally valid for lensed signals in doubly 
imaged systems (doubles). 
The lensed signals studied here have $|\mu|\sim[0.5, 10]$ and 
$\Sigma_\bullet\sim[10, 10^3]~{\rm M}_\odot/{\rm pc^2}$. 
We find that the microlensing effects are more sensitive to the 
macro-magnification than the underlying microlens density, even if the 
latter exceeds~$10^3~{\rm M}_\odot/{\rm pc^2}$. 
The mismatch between lensed and unlensed GW signals rarely exceeds 1\% 
for nearly all binary black hole sources in the total mass range
[10~M$_\odot$,~200~M$_\odot$]. 
This implies that microlensing is not expected to affect the detection 
or the parameter estimation of such signals and does not pose any 
further challenges in identifying the different lensed counterparts 
when macro-magnification is~${\leq}10$. 
Such a magnification cut is expected to be satisfied by~${\sim}50\%$ 
of the detectable pairs in quads and~${\sim}90\%$ of the doubles in 
the fourth observing run of the LIGO--Virgo detector network. 
\end{abstract}
%%%%%%%%%%%%%%%%%%%%%%%%%%%%%%%%%%%%%%%%%%%%%%%%%%%%%%%%%%%%

%%%%%%%%%%%%%%%%%%%%%%%%%%%%%%%%%%%%%%%%%%%%%%%%%%%%%%%%%%%%
%%%%%%%%%%%%%%%%%%%%%%%%%%%%%%%%%%%%%%%%%%%%%%%%%%%%%%%%%%%%

\section{Introduction}
\label{sec:intro}

The Advanced Laser Interferometer Gravitational-wave Observatory 
\citep[LIGO;][]{2015CQGra..32g4001L}, 
the Advanced Virgo \citep{2015CQGra..32b4001A},
and the Kamioka Gravitational Wave Detector \citep[KAGRA;][]{2021PTEP.2021eA101A} 
have detected a total of 90 gravitational wave (GW) signals coming from 
merging binaries at cosmological distances~\citep{2021arXiv211103606T}.
With an increase in the sensitivity of the current \citep{2018LRR....21....3A} 
and future detectors , such as the 
Cosmic Explorer~\citep[CE:][]{2021arXiv210909882E},
the Einstein telescope~\citep[ET:][]{2020JCAP...03..050M},
the Deci-hertz Interferometer Gravitational wave Observatory~\citep[DECIGO;][]{2021PTEP.2021eA105K},
and the Laser Interferometer Space Antenna~\citep[LISA:][]{2020GReGr..52...81B}, 
the number of detectable GW events is expected to increase considerably
\citep[e.g.,][]{2018LRR....21....3A, 2021NatRP...3..344B}.

Like electromagnetic (EM) radiation, GWs also get affected by the 
mass distribution around the path between the GW source and the observer 
through the phenomenon of gravitational lensing~\citep[e.g.,][]{1971NCimB...6..225L, 1974IJTP....9..425O}.
In a typical case of gravitational lensing of the EM radiation (for 
example, lensing due to a stellar mass object or a galaxy), the geometric 
optics approximation is sufficient to study the lensing effects as the 
wavelength of the EM radiation is much smaller compared to the Schwarzschild 
radius corresponding to the lens mass ($\lambda~{\ll}~R_{\rm Sch}$).
However, if the wavelength of the EM radiation is of the order of the 
Schwarzschild radius of the lens 
mass~($ \lambda~{\sim}~\mathcal{O}(R_{\rm Sch})$), then the geometrical 
optics approximation is not adequate and one needs to consider the 
corrections coming from the wave optics~\citep[e.g.,][]{1995ApJ...442...67U}.
The same is also applicable to the gravitational lensing of GWs.
For GWs detected by the LIGO in the frequency band [$10,10^4$]~Hz, if 
the lens is a galaxy or a galaxy cluster, the geometric optics approximation 
works well.
However, for the lens masses in the mass range $[10,10^4]{\rm M}_\odot$, 
the GW wavelength (in the LIGO frequency range) is of the order of 
the Schwarzschild radius corresponding to the lens mass (see figure 1 
of~\citealt{2020MNRAS.492.1127M}).
This results into non-negligible wave optics effects and introduces 
frequency-dependent modulations in the GW signal~\citep[e.g.,][]{1981Ap&SS..78..199B, 
1986PhRvD..34.1708D,1998PhRvL..80.1138N, 1999PhRvD..59h3001B, 
1999PThPS.133..137N, 2003ApJ...595.1039T,2018PhRvD..98j3022C}.

In the strong lensing regime, when a GW signal is lensed by a 
galaxy-scale lens, it gets (de-)amplified in an achromatic fashion by 
a constant factor, $\sqrt{|\mu|}$, where $|\mu|$ is the strong lens 
(macro-) magnification.
This can increase or decrease the signal-to-noise (SNR) ratio of the GW signal.
As the GW signal amplitude is proportional to the chirp mass of the 
binary source and inversely proportional to the source luminosity distance, 
strong lensing can bias the measurement of these parameters and lead to 
erroneous results\citep[e.g.,][]{2018arXiv180205273B} for individual 
signals and can also affect the statistical 
inferences~\citep[e.g.,][]{2018MNRAS.480.3842O}.
The de-amplification of one of the strongly lensed signal can also result 
into sub-threshold events~\citep[e.g.,][]{2019arXiv190406020L, 2020PhRvD.102h4031M}.
Apart from a frequency-invariant (de-)amplification, strong lensing also 
introduces a constant phase shift in the strongly lensed signal. 
For minima, saddle points, and maxima, the phase shifts are 0, $\pi/2$, 
and $\pi$, respectively~\citep[e.g.,][]{2017arXiv170204724D}.
Such a phase shift can help us identify different counterparts of the 
strongly lensed signal \citep[e.g.,][]{2020arXiv200712709D, 
2021arXiv210409339L, Janquart_2021, Vijaykumar2022}.

A typical galaxy-scale lens also harbors small compact objects (e.g., 
stars, stellar remnants, possible compact dark matter objects).
As mentioned above, these point-like objects can lead to frequency-dependent 
effects in the strongly lensed GW signal in the LIGO frequency band 
which can affect the measurement of parameters other than the luminosity 
distance and the chirp mass~\citep[e.g.,][]{2018PhRvD..98j3022C, 
2018PhRvD..98h3005L, 2020MNRAS.492.1127M}.
For an isolated microlens, the amplitude of the wave effects depends 
only on the lens mass and the impact 
parameter~\citep[e.g.,][]{2003ApJ...595.1039T, 2021arXiv211210773B}.
However, lensing by an isolated point mass is not suitable for the 
microlensing studies of strongly lensed GW signals.
Thus, we consider the more plausible physical scenario of a GW source 
strongly lensed by a galaxy and further microlensed by a population 
of microlenses belonging to the galaxy. 
In such cases, we need to take into account the strong lensing 
macro-magnification and the properties of the microlens population.
The effect of microlens population on the strongly lensed GW signals 
in the high-magnification regime has been studied in~\citet{2019A&A...627A.130D} 
although their underlying microlens surface mass densities were similar 
to those found in the intra-cluster medium rather than galaxy-scale lenses.
In \citet{2021MNRAS.508.4869M}, we studied the effect of microlens 
population (appropriate for galaxy-scale lenses) considering specific 
scenarios for individual signals to identify the parameters which 
govern the microlensing effects.

In this work, we study the galaxy-scale lens population as a whole to 
determine the fraction of lenses that could be affected by microlensing 
and to understand the extent of these effects.
As the GW source is effectively a point source, strong lensing of 
GWs can, in principle, lead to highly-magnified GW signals.
In such a scenario, we expect to observe significant microlensing effects.
However, the probability of a source being highly magnified decreases
with an increase in macro-magnification. 
This is because the total area in the source plane with magnification 
above a threshold decreases with an increase in the 
threshold~\citep[e.g.,][]{2019A&A...625A..84D}. 
Hence, the fraction of lensed GW systems with high magnification will be small. 
According to our projections, only ${\sim}10\%$  of the doubly lensed 
systems (doubles) with SNR$\geq8$ in the fourth observing run of 
LIGO--Virgo detector network will have macro-magnification~${\geq}10$. 
On the other hand, the quadruply lensed systems (quads) tend to have 
lensed images with higher magnifications and thus, ${\sim}50\%$ of 
the brightest pair will have macro-magnification~$\geq10$.
In addition, studying the frequency-dependent effects in the lensed GW 
signals with high macro-magnification becomes computationally expensive 
as we need to simulate a bigger patch (with large number of microlenses) 
in the lens plane.
Hence, in this work, we only focus on lensed GW systems where all 
of the lensed signals have macro-magnification ($|\mu|$)~${\leq}10$.

We simulate quads assuming the singular isothermal 
ellipsoid~\citep[SIE;][]{1994A&A...284..285K} density profile for the 
galaxy-scale lens.
Considering the fact that in quads the central (de-magnified) image 
(image nearest to the lens center) can have very high microlens density 
values becoming computationally very expensive, we apply an upper-limit of 
$10^3\:{\rm M}_\odot/{\rm pc^2}$ on the microlens density.
Therefore, our main lens sample comprises of lensed images formed with 
microlens densities of $\Sigma_\bullet<10^3\:{\rm M}_\odot/{\rm pc^2}$.
Nevertheless, we also study the microlensing effects in four individual 
images with $\Sigma_\bullet\sim[10^3, 10^4]\:{\rm M}_\odot/{\rm pc^2}$.
To quantify the effect of microlensing in a lensed signal, we calculate 
the mismatch between the lensed and unlensed GW signal. 
We assess the applicability of our results from the analysis of quads 
sample to those from doubles.
Our analyses allow us to generalize our results for all lensed signals 
with macro-magnification ${\leq}~10$. To our knowledge, this is the 
first detailed study of microlenisng effects in different counterparts 
of a lensed GW system.

As mentioned earlier, in a galaxy-scale lens system, the multiple 
lensed images have different underlying macro-magnifications and 
microlens densities. 
This leads to the possibility that the counterparts of a strongly lensed 
GW signal have different microlensing features and one may not be able 
to identify that the lensed GW signals are coming from the same source 
as pointed out in \citet{2020MNRAS.492.1127M}. 
Hence, it is important to study how microlensing affects the common 
origin hypothesis of different lensed counterparts. 
Since we are simulating all of the lensed images formed in a given lens 
system, we can address this question appropriately by comparing the 
microlensing effects in different lensed counterparts and the 
corresponding (mis)match values.

This paper is organized as follows.
In Section~\ref{sec:lens_systems}, we briefly discuss the method used 
to the simulate strong lens systems and the corresponding lensed image 
properties.
Here we also describe the method used to generate the microlens population.
In Section~\ref{sec:ml_effects}, we study the effect of microlens 
population on different counterparts of strongly lensed GW signal in 
the quadruply lens systems.
In Section~\ref{sec:mismatch}, we study the effect of the microlens 
population on the mismatch between the unlensed and the lensed signals.
In Section~\ref{sec:high_stelden}, we study the faintest fourth image 
of the quad formed in high stellar density environments 
(${>}10^3\:{\rm M}_\odot/{\rm pc^2}$).
In Section~\ref{sec:two_image}, we discuss the applicability of our 
results in doubly lensed systems.
Discussion and conclusions are given in Section~\ref{sec:conclusions}. 
We also discuss the future work in this section.
The cosmological parameters used in this work to calculate the various 
quantities are: 
$H_0 {=} 67.27 \: {\rm km} \: {\rm s}^{-1} \: {\rm Mpc}^{-1}$, 
$\Omega_\Lambda {=} 0.69$, and  $\Omega_m {=} 0.31$.

%%%%%%%%%%%%%%%%%%%%%%%%%%%%%%%%%%%%%%%%%%%%%%%%%%%%%%%%%%%%%%%%%%%%%%%%%%%%%%%%%%%%%%%%%%%
\begin{figure}
    \centering
    \hspace{-0.5cm}\includegraphics[height=9cm, width=9cm]{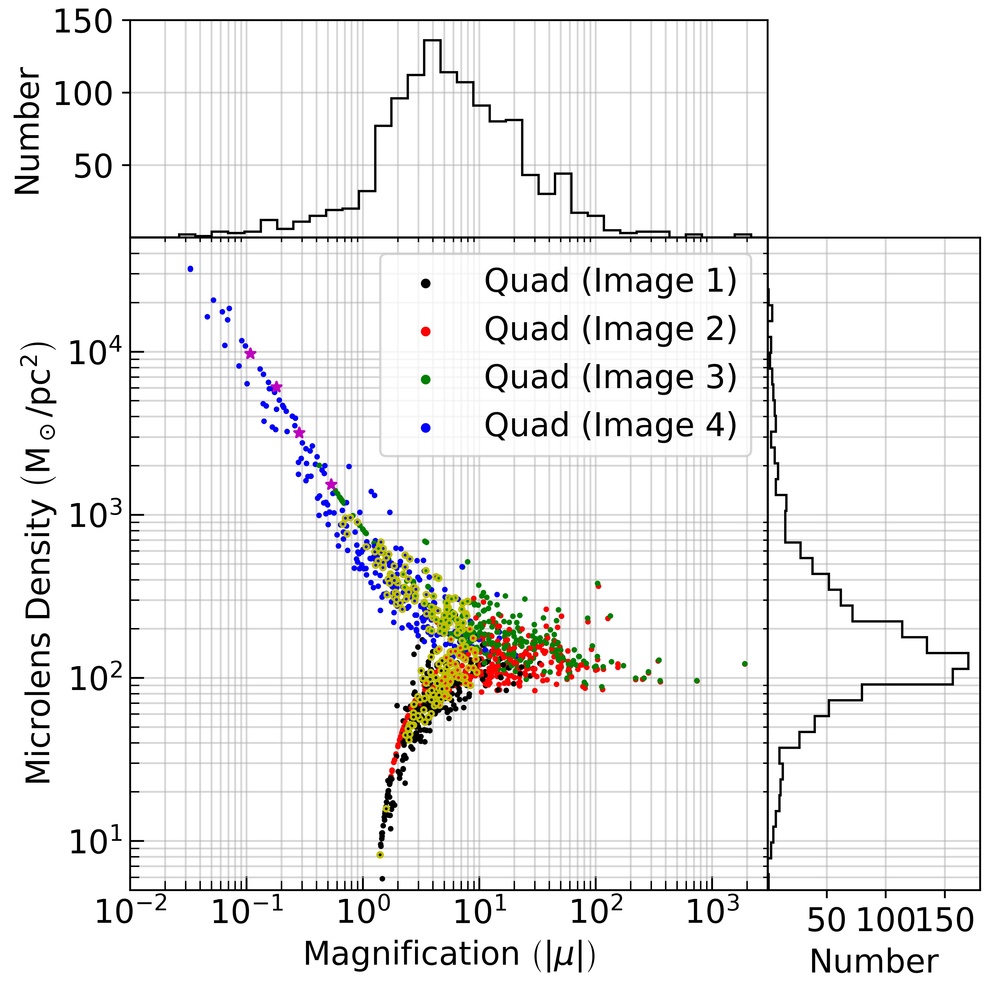}
    \caption{Distribution of images formed in quad lens systems as a function of 
    macro-magnification ($|\mu|$) and microlens density ($\Sigma_\bullet$).
    The black, red, green, and blue points represent the \emph{image-1} (minima), 
    \emph{image-2} (minima), \emph{image-3} (saddle points) and \emph{image-4} 
    (saddle points) which are ordered according to their time-delay compared to the global
    minimum, i.e., \emph{image-1}. The \emph{yellow-edge} points represent the 50 lens 
    systems which are analysed in this work whereas the filled maroon stars represent 
    additional 4 cases studied for the saddle points corresponding to \emph{image-4} as 
    they are located in regions with extreme microlens densities
    (${>}10^3~{\rm M}\odot/{\rm pc^2}$) albeit at low magnifications. We also show the 1D
    histograms of the magnification (top) and microlens density (right) for all four images 
    combined.}
    \label{fig:sl_systems}
\end{figure}
%%%%%%%%%%%%%%%%%%%%%%%%%%%%%%%%%%%%%%%%%%%%%%%%%%%%%%%%%%%%%%%%%%%%%%%%%%%%%%%%%%%%%%%%%%%

\section{Simulating lens systems}
\label{sec:lens_systems}

In this section, we briefly discuss the methodology used to simulate the strong lens 
systems and the corresponding microlens population for individual strongly lensed images.

%%%%%%%%%%%%%%%%%%%%%%%%%%%%%%%%%%%%%%%%%%%%%%%%%%%%%%%%%%%%%%%%%%%%%%%%%%%%%%%%%%%%%%%%%%%
\begin{figure*}
    \centering
    \includegraphics[width=16cm, height=4cm]{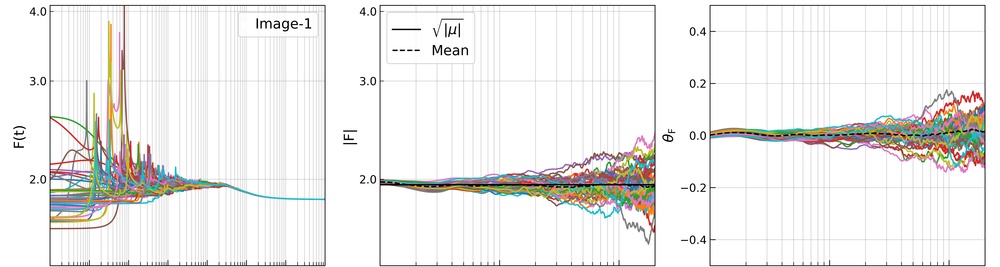}
    \includegraphics[width=16cm, height=4cm]{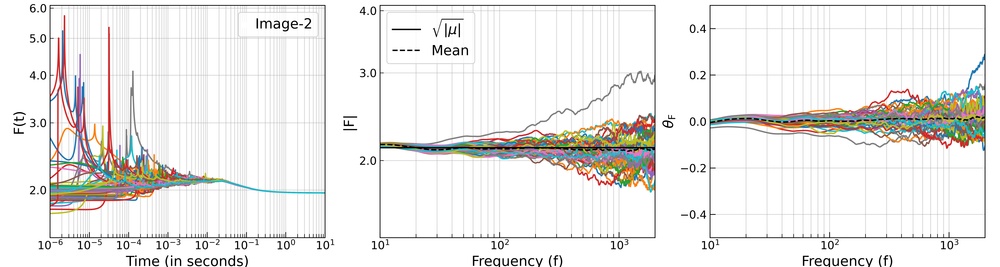}
    
    \vspace{0.5cm}
    
    \includegraphics[width=16cm, height=4cm]{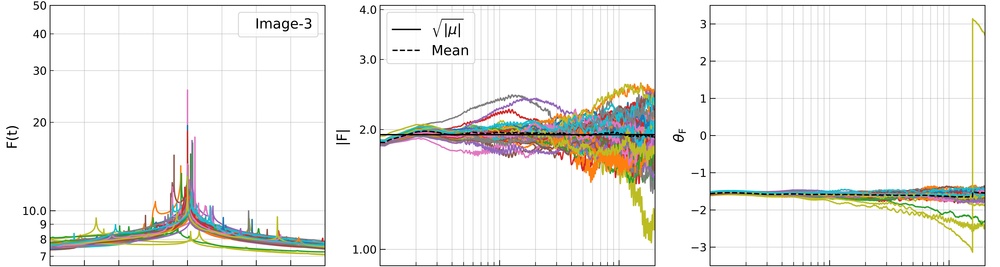}
    \includegraphics[width=16cm, height=4cm]{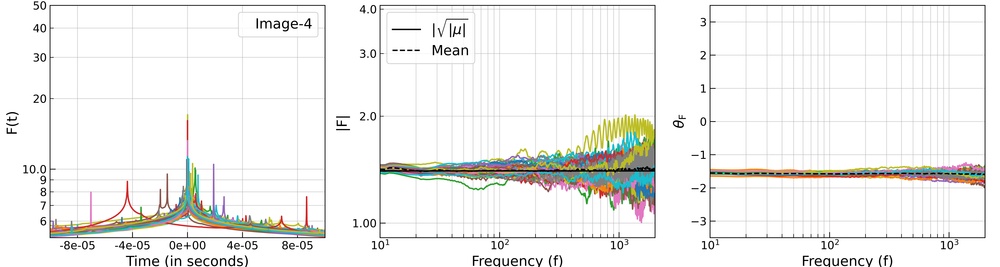}
    \caption{Microlensing effects in all four images (\emph{image-1, 2, 3, 4}) of one quad 
    lens system. From top-to-bottom, each row corresponds to one of the images in the quad 
    (as indicated by the labels) and each panel shows 50 curves corresponding to the 50 
    realizations for each image. The left, middle, and right columns represent the $F(t)$, 
    the absolute value of amplification factor ($|F|$), and phase of the amplification 
    factor ($\theta_{\rm F}$).  The $F(t)$ curves contain discontinuities and spikes which 
    correspond to micro-minima and micro-saddle points, respectively. In the middle column, 
    the black solid line represents the amplification factor, $\sqrt{|\mu|}$, in the 
    geometric-optics limit which is independent of the frequency. The black dashed curves 
    (in middle and right columns) represent the mean over the 50 realizations.}
    \label{fig:ml_system}
\end{figure*}
%%%%%%%%%%%%%%%%%%%%%%%%%%%%%%%%%%%%%%%%%%%%%%%%%%%%%%%%%%%%%%%%%%%%%%%%%%%%%%%%%%%%%%%%%%%

%%%%%%%%%%%%%%%%%%%%%%%%%%%%%%%%%%%%%%%%%%%%%%%%%%%%%%%%%%%%%%%%%%%%%%%%%%%%%%%%%%%%%%%%%%%
\begin{figure*}
    \centering
    \includegraphics[width=8cm, height=12cm]{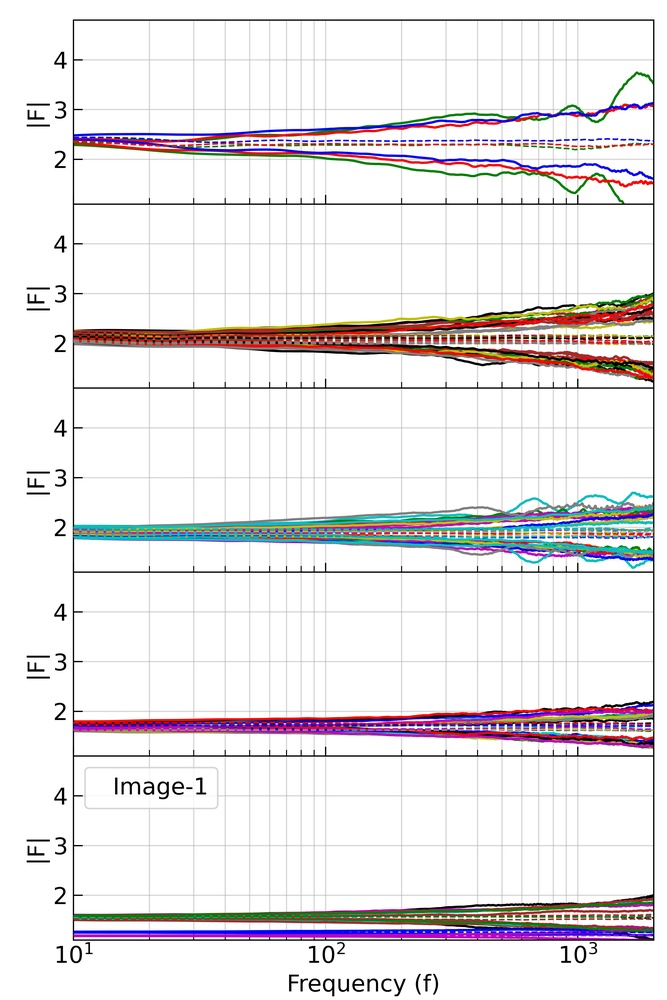}
    \includegraphics[width=8cm, height=12cm]{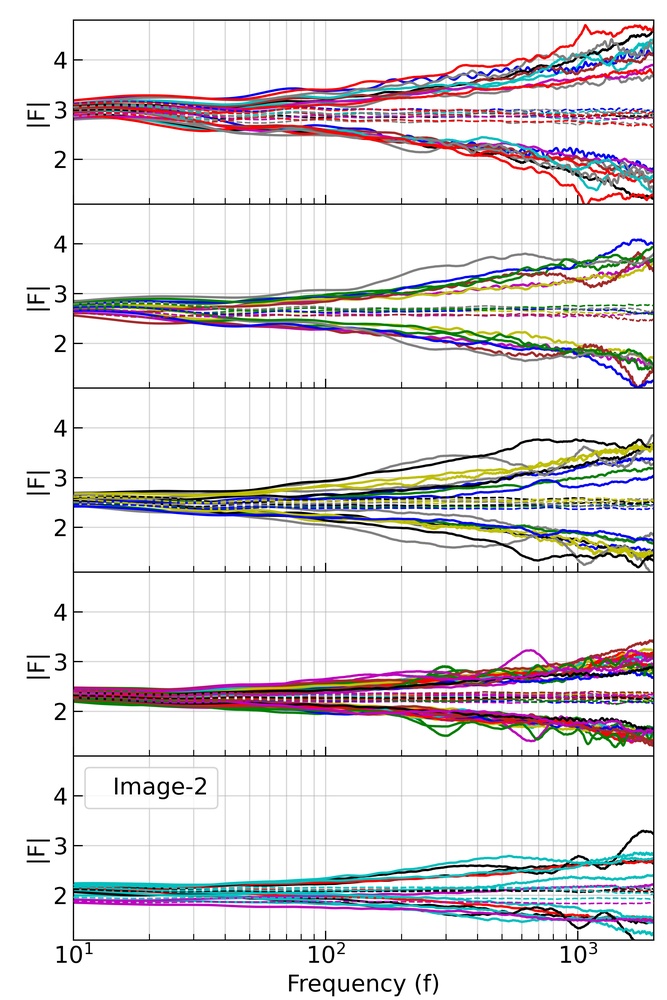}
    \caption{Microlensing effects in minima corresponding to \emph{image-1} (left) and 
    \emph{image-2} (right). The dotted lines represent the frequency-dependent mean of the 
    absolute amplification factor ($|F|$) based on 50 realizations. The solid lines represent 
    the $3-\sigma$ scatter around the mean. From the bottom-to-top, different lens systems are 
    grouped and split over multiple panels according to their macro-magnifications
    ($\sqrt{|\mu|}$). Doing so gives improved visual clarity in order to appreciate the
    magnification-dependence of microlensing effects.}
    \label{fig:ml_system_minima}
\end{figure*}
%%%%%%%%%%%%%%%%%%%%%%%%%%%%%%%%%%%%%%%%%%%%%%%%%%%%%%%%%%%%%%%%%%%%%%%%%%%%%%%%%%%%%%%%%%%

%%%%%%%%%%%%%%%%%%%%%%%%%%%%%%%%%%%%%%%%%%%%%%%%%%%%%%%%%%%%%%%%%%%%%%%%%%%%%%%%%%%%%%%%%%%
\begin{figure*}
    \centering
    \includegraphics[width=8cm, height=12cm]{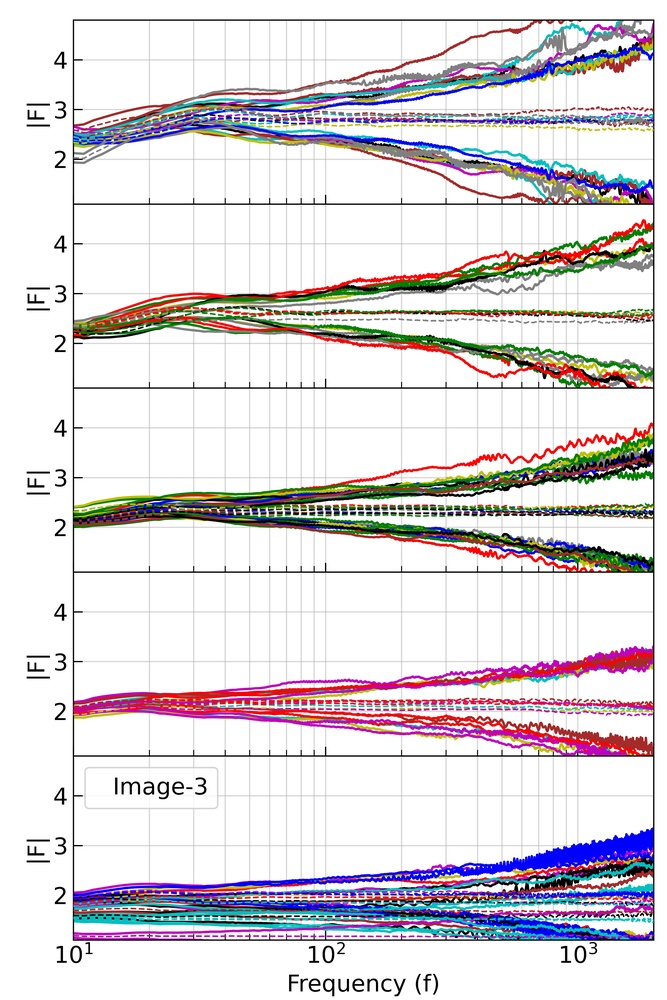}
    \includegraphics[width=8cm, height=12cm]{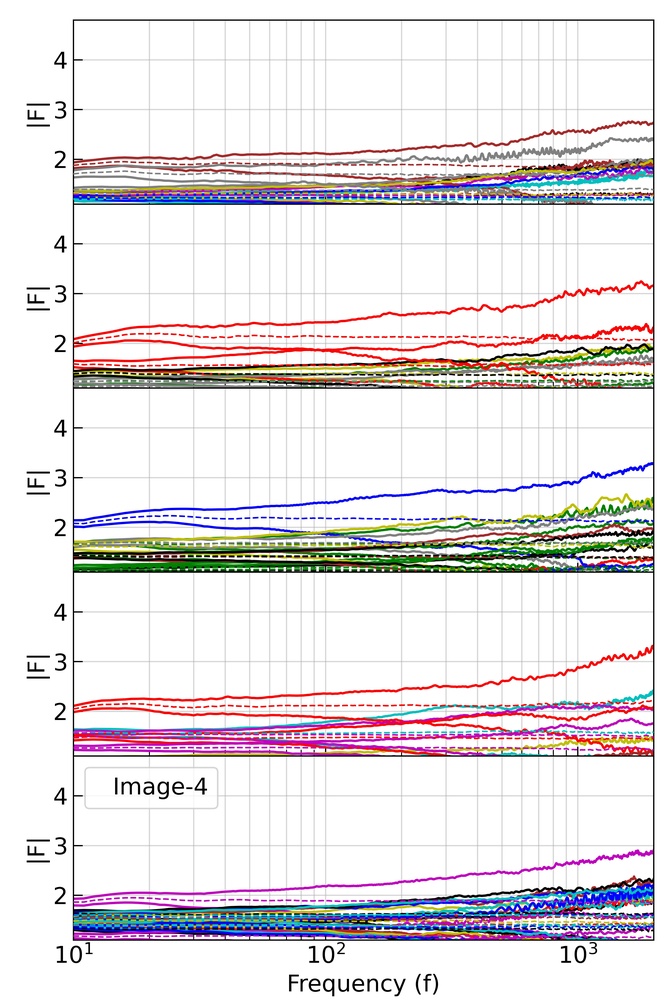}
    \caption{Same as Figure~\ref{fig:ml_system_minima} except that the microlensing effects 
    are shown for saddle points corresponding to \emph{image-3} (left) and \emph{image-4}
    (right).}
    \label{fig:ml_system_saddle}
\end{figure*}
%%%%%%%%%%%%%%%%%%%%%%%%%%%%%%%%%%%%%%%%%%%%%%%%%%%%%%%%%%%%%%%%%%%%%%%%%%%%%%%%%%%%%%%%%%%

\subsection{Simulating strongly lensed systems}
\label{ssec:sl_simulate}

A strong lens system contains three parts: the source, the lens, and the observer (us).
In geometric optics (under thin lens approximation), %the lens
gravitational lens equation
gives a mapping between source plane and lens (image) plane 
\citep[e.g.,][]{1992grle.book.....S, 1996astro.ph..6001N}.
As we are mainly focusing here on the lensing of gravitational waves
due to a galaxy scale lens in the LIGO frequency range, a typical source will be a merging 
binary black hole (BBH) system and the lens is assumed to be an elliptical galaxy following 
a singular isothermal ellipsoid \citep[SIE;][]{1994A&A...284..285K} density profile.

We use the realistically generated mock sample of strong lens systems from
\citet[][hereafter MM21]{2021arXiv211103091M}. We briefly describe below the different 
assumptions and ingredients of the framework. 
Assuming that the BBH source is spinless with zero eccentricity, the source population can 
be described by modeling the three quantities: 
(i) BBH merger rate, 
(ii) BBH source redshift, and 
(iii) BBH component masses.
The BBH merger rate has been approximated using the analytical formula given in 
equation 11 of MM21 (taken from \citealt{2018MNRAS.480.3842O}).
Such a functional form roughly reproduce the numerical results shown  in
\citet{2016MNRAS.456.1093K} and \citet{2017MNRAS.471.4702B}.
The source redshift distribution (in the observer frame) %is given by equation which 
depends on the the merger rate and the comoving volume (See equation 12 in MM21). 
Finally, the BBH component masses are drawn using the \textsc{power law + peak} 
model given in \citet{2021PhRvX..11b1053A}.

The lens galaxy population is defined via the:
(i) redshift, 
(ii) velocity dispersion,  and 
(iii) ellipticity (magnitude and a position angle), of the lens.
The lens redshift is drawn uniformly in the comoving volume between observer 
and source. The velocity dispersion is drawn from the modified Schechter function, an 
observational fit to the Sloan Digital Sky Survey (SDSS) data,  following \citep{2007ApJ...658..884C}.
The lens ellipticity is drawn from the observed distribution of the 
elliptical galaxies in the SDSS data \citep[see figure 4 in ][]{2008MNRAS.388.1321P}.

An SIE lens model gives rise to formation of either two or four image geometry known
as doubles and quads \citep{1994A&A...284..285K}.
In doubles, the pair of images are of different types i.e., a minimum and a saddle point 
whereas in quads, two of the images are minima and the other two are saddle points.
In general, the lensed images are (de-)magnified by different factors and the signal from 
them takes different amount of time to reach the observer.
Typically, the time delay between these lensed images ranges from hours to months for 
galaxy-scale lenses.
The macro-magnification factor corresponding to these different images also depends on 
the source size: smaller the source, higher the magnification.
As mentioned above, since the BBH merger is effectively a point source, the corresponding 
strongly lensed GW signals can achieve very high magnification 
\citep[][]{2019A&A...627A.130D, 2021MNRAS.508.4869M}.
To verify that a GW signal is strongly lensed, we need to observe at least two of 
the strongly lensed counterparts. 
Hence, we consider lens systems wherein the second brightest lensed GW signal has a
three-detector-network SNR of 8 or above. 
We note that this detectability criteria is arbitrary and does not affect the 
conclusions of our work as explained in Section~\ref{sec:ml_effects}. 
Following the aforementioned procedure, we draw a sample of 300 quads and 200 doubles. 
This sample contains systems with individual image magnification values up to $10^3$.
However, as we  are mainly focussing on macro-magnification ($|\mu|$) ${\leq}10$ 
regime we draw a sub-sample of 50 quad systems (randomly) such that all lensed images
satisfy $\{|\mu|, \Sigma_\bullet\} \leq \{10, 10^3\:{\rm M}_\odot/{\rm pc}^2\}$.

\subsection{Simulating microlens population}
\label{ssec:ml_simulate}

Once we obtain the strong lens systems, the next step is to determine the microlens 
densities at the position of the strongly lensed images. 
The microlens density at an image position consists of stars, stellar remnants 
(white dwarf, neutron star and black hole) and possible compact dark matter objects
like primordial black holes \citep[PBH; e.g.,][]{2016PhRvL.116t1301B, 2017PhRvL.119m1301K}.
In this work, we assume that all of the dark matter is in the form of a smooth component
leaving only stars and stellar remnants as possible microlenses.
To estimate the projected stellar surface mass density at the image positions, 
we use the S\'ersic profile following \citet[][see equation 8]{2019MNRAS.483.5583V}.
Since, elliptical galaxies have little to no star formation, only the stars with masses 
${\lesssim}1.5{\rm M}_\odot$ will contribute to the corresponding projected stellar 
density as massive stars (${\gtrsim}1.5{\rm M}_\odot$) will complete their life in ${\sim}1.5$Gyr.
Hence, we draw stellar masses in only the mass range $[0.08, 1.5]{\rm M}_\odot$ using
the Salpeter initial mass function \citep[IMF,][]{1955ApJ...121..161S}.
To sample the remnant population, we estimate the corresponding projected surface mass 
density using the initial-final mass function from the Binary 
Population And Spectral Synthesis \citep[BPASS;][]{2017PASA...34...58E}.
Depending on the IMF and the metallicity, the fraction of remnant mass to the stellar 
mass can be from ${\sim}5\%$ to ${\sim}20\%$.
In our work, we assume this fraction to be $10\%$ at the image positions.
As the remnants are dark, we need to add the corresponding surface density to the
stellar surface mass density to get final microlens surface mass density.
Hence, our final microlens density at an image position is sum of the stellar density 
and remnant density.

The calculated projected microlens density ($\Sigma_{\rm \bullet}$) for different images 
and the corresponding macro-magnification for our simulated quads are shown in
Figure~\ref{fig:sl_systems}.
The black, red, green, and blue points represent the \emph{image-1, 2, 3, 4} of 
the quad systems, respectively.
The numbering of the images follows the same order as their arrival times, that is,  
signal from \emph{image-1} is seen first whereas signal from \emph{image-4} is 
seen at the end.
Here, we only plot the unsigned magnifications ($|\mu|$) for all of the images 
leading to a \emph{butterfly}-shaped distribution in the $|\mu|$--$\Sigma_\bullet$ plane.
We note that the \emph{butterfly} shape is not dependent on the detection threshold 
criteria applied in Section~\ref{ssec:sl_simulate}
As expected the magnification for minima (\emph{image-1} and \emph{image-2}) 
always have $\mu{\gtrsim}1$ whereas the saddle points in addition can also have 
$|\mu|{<}1$.
Apart from that, the microlens densities for \emph{image-1, 2, 3, 4} are in an 
increasing order.
This trend is expected since the saddle points form closer to the center of the 
lens galaxy in regions of high densities as compared to the minima.
In addition, only \emph{image-2} and \emph{image-3} attains $|\mu|{>}20$ simultaneously 
as they tend to form in a pair near the critical curve in the lens plane when 
the source lies near the caustic in the source plane.
The 1D histograms of the magnification ($|\mu|$) and the microlens density ($\Sigma_\bullet$) 
are also shown on the panels at the top and to the right, respectively.
The magnification peaks below 10 and thus implying most of the individual lensed 
signals~(${\sim}75\%$) in our sample have macro-magnification values below~${\leq}10$.
However, in our sample, nearly~${\sim}36\%$ and~${\sim}90\%$ quad and double lens 
systems, respectively, have their brightest image with macro-magnification~${\leq}10$.
On individual image basis in a given sample of quad lens systems, ${\sim}94\%$,  
${\sim}36\%$,  ${\sim}40\%$, and ${\sim}96\%$ of \emph{image-1}, \emph{image-2}, 
\emph{image-3}, \emph{image-4}, respectively, are expected to satisfy the above 
magnification cut.
On the other hand, the microlens density peaks around ${\sim}100~{\rm M}_\odot/{\rm pc}^2$
reflecting the typical microlens densities at the image position in the galaxy-scale lenses.
The yellow-edge circles mark the lensed images corresponding to 50 lens 
systems considered in our analysis.
The solid maroon stars show the four images corresponding to 
\emph{image-4} that are chosen to study the high density regions, specifically 
(see Section~\ref{sec:high_stelden}).

\section{Effect of microlens population in strongly lensed systems}
\label{sec:ml_effects}

To study the effect of microlens population on the strongly lensed GW signal, we need to 
compute the amplification factor, $F(f)$, which quantifies the wave effects. 
In the context of GW, it represents the ratio of lensed and unlensed GW waveform 
in the frequency domain.
We consider a patch of $2{\rm pc}\times2{\rm pc}$ in the lens plane and randomly distribute
microlenses in it which are drawn using the procedure discussed in~\ref{ssec:ml_simulate}.
Such a patch size ensures that we can go up to a time delay of the order of $0.1$ seconds in 
the lens plane and, consequently, cover the LIGO frequency range while computing $F(f)$.
In the absence of microlensing, there is only one image corresponding to the macro-image 
forming at the center of the patch.
Introducing microlenses divide this macro-image into multiple micro-images.
To calculate the $F(f)$, we first calculate its Fourier transform, $F(t)$, in time-domain.
The values of $F(t)$ depends on the time delay of different micro-images forming in the patch.
In principle, the spread of micro-images in the lens plane depends on the macro-magnification~($|\mu|$)
but as we are mainly focussing on~$|\mu|<10$ regime, the above patch size is enough.
The value of macro-magnification along with the microlens density also determines the underlying 
computation time: less the macro-magnification or microlens density, less the computation time.
We refer readers to~\citet{1995ApJ...442...67U} and~\citet{2021MNRAS.508.4869M} for more 
details about the computation method.

Following the above described method, we generate $F(t)$ and $F(f)$ corresponding to all 
realizations in all 50 lensed systems.
In general, the frequency-dependent microlensing effects will be different for different 
strongly lensed signals as the underlying macro-magnification ($|\mu|$) and microlens 
densities~($\Sigma_\bullet$) vary. 
However, for a better visualization and understanding of microlensing effects across the 
different images of a system, we show the microlensing effects for all 
of the four lensed images of one lens system in Figure~\ref{fig:ml_system}.
The top two rows represent the macro-minima (\emph{image-1} and \emph{image-2}) 
whereas the bottom two rows represent the macro-saddle points (\emph{image-3} and 
\emph{image-4}).
From left to right, we show the Fourier transform of the amplification factor, $F(t)$, 
the absolute value of the amplification factor, $|F(f)|$, and the phase of the amplification 
factor, $\theta_{\rm F} =-\iota \ln\left(F/|F|\right)$, respectively.
Each panel shows 50 curves corresponding to the 50 realizations for each image.
The $F(t)$ curves start from zero for minima (\emph{image-1} and \emph{image-2}) whereas 
for the saddle-point, the $F(t)$ curves have non-zero values at negative times.
This happens because when a macro-minimum is microlensed,  we can always find a global minimum
and calculate the time delay of other micro-images with respect to this global
minimum \citep[e.g.,][]{2011MNRAS.411.1671S}.
On the other hand, a macro-saddle point has a global saddle and other micro-images
can have both positive and negative time delays with respect to the global saddle.
In each $F(t)$ curve, we can see two different kind of features, discontinuities and spikes
representing the minima and saddle-point micro-images in the lens plane, respectively.
In general, a micro-image forming at a time delay $\delta t$ (compared to the global minimum 
or global saddle-point) effects $F(f)$ at $f\gtrsim (\delta t)^{-1}$. 
Therefore,  micro-images forming at lower (higher) time-delays lead to slow (rapid) variations in amplification factor, $F(f)$, curve 
at low frequencies, 10 -- 100 Hz.

In the middle column of Figure~\ref{fig:ml_system}, we show the mean of the 50 realizations 
(dashed-black curve) and the amplification factor in geometric optics 
($\sqrt{|\mu|}$, solid black curve) which is independent of the GW frequency.
We find that the mean $|F|$ is close to $\sqrt{|\mu|}$ at all frequencies. 
The same is also true for the phase in the right column where the mean phase
for minima and saddle points remain very close to their geometrical optics values,
0 and $\pi/2$, respectively \citep{2017arXiv170204724D}.
This implies that, although for this particular lens system, microlensing introduces additional 
frequency-dependent fluctuations in the $|F|$ and $\theta_{\rm F}$, these fluctuations are unique 
to each realization otherwise we would have observed coherent features in the mean value.
In addition, the amplitude of these oscillations increases as we go towards higher 
GW frequencies owing to the multiple faint micro-images forming at higher time-delay values.
For \emph{image-3}, the yellow phase curve, for one of the realizations, shows a discontinuity 
from $-\pi$ to $\pi$ around 1500~Hz which is an artifact arising from the choice of the limits
used to specify the phase.

%%%%%%%%%%%%%%%%%%%%%%%%%%%%%%%%%%%%%%%%%%%%%%%%%%%%%%%%%%%%%%%%%%%%%%%%%%%%%%%%%%%%%%%%%%%
\begin{figure*}
    \centering
    \includegraphics[width=8.8cm, height=6.5cm]{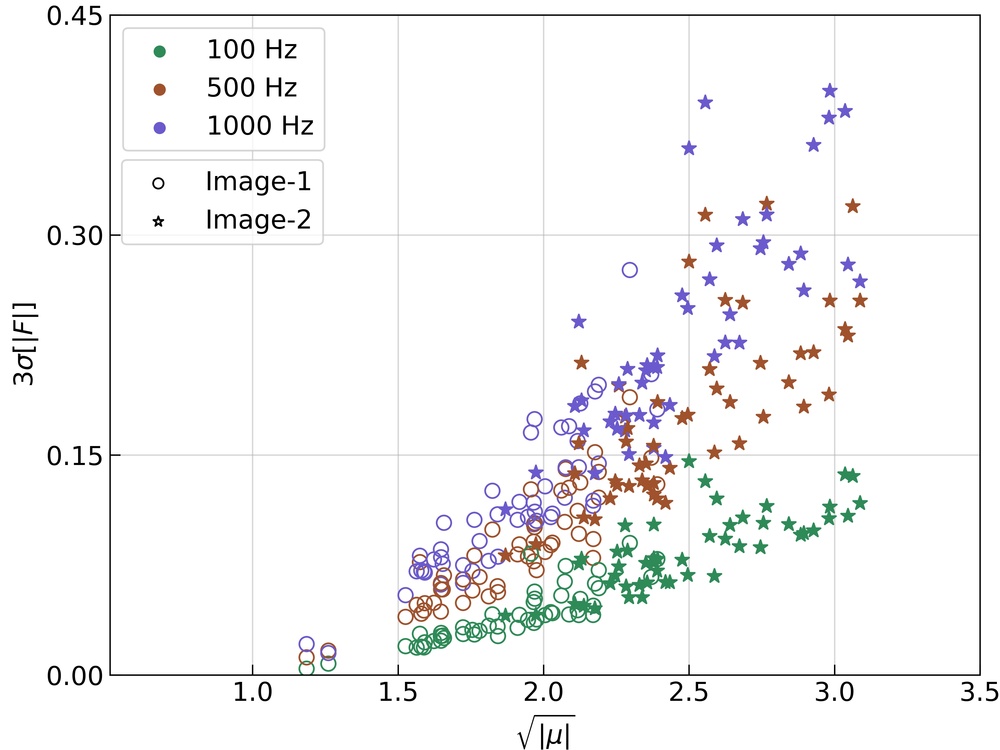}
    \includegraphics[width=8.8cm, height=6.5cm]{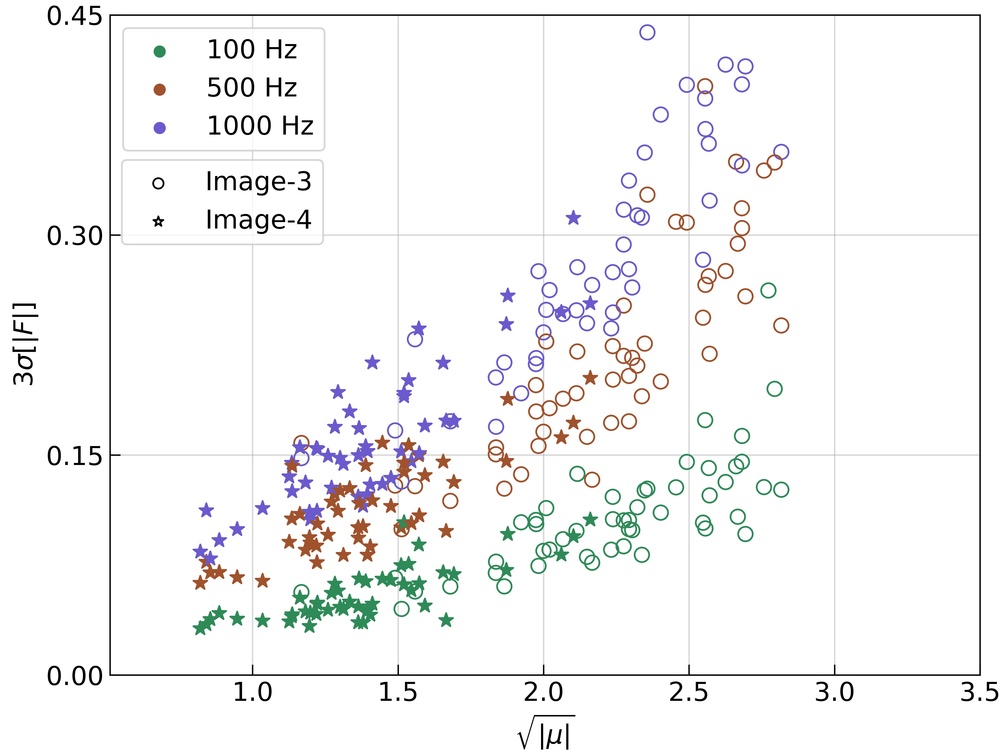}
    \caption{Scatter as a function of the geometric-optics amplification factor 
    ($\sqrt{|\mu|}$) at different GW frequencies, namely, 100~Hz, 500~Hz, and 1000~Hz. 
    The open circles and filled-stars represent \emph{image-1} and \emph{image-2} 
    in the left panel and \emph{image-3} and \emph{image-4} in right panel, respectively. 
    At 100 Hz, the $3\sigma$ values increase (roughly) linearly with increasing 
    $\sqrt{|\mu|}$. However, as we increase the frequency (500 Hz and 1000 Hz), the 
    $3\sigma$ values increase more rapidly with $\sqrt{|\mu|}$ and show more 
    scatter compared to 100 Hz implying stronger dependency on the strong lensing 
    amplification factor ($\sqrt{|\mu|}$). See Section~\ref{sec:ml_effects} for more details.}
    \label{fig:ml_linear}
\end{figure*}
%%%%%%%%%%%%%%%%%%%%%%%%%%%%%%%%%%%%%%%%%%%%%%%%%%%%%%%%%%%%%%%%%%%%%%%%%%%%%%%%%%%%%%%%%%%

In Figures~\ref{fig:ml_system_minima} and \ref{fig:ml_system_saddle}, we show the mean and scatter 
across different realizations for each minimum and saddle point, respectively, from all of the 50 systems.
In these figures, the mean $|F|$ (dashed curves) and a ${\pm}3\sigma$ (i.e., ${\pm}3\times$ 
standard deviation) region around the mean (solid curves) of a given color represent a lens system. 
In each column, the different lensed images are grouped according to their $\sqrt{|\mu|}$ and 
plotted in an increasing order from bottom-to-top in different rows for improved clarity.
In \emph{image-1, 4}, we observe that the mean curves do not show significant 
fluctuations and remain approximately constant in the frequency range [10, 2000]~Hz.
On the other hand, in \emph{image-2, 3} at macro-magnification~${\sim}10$, we begin to see 
some coherent deviations (compared to the geometric optics value, $\sqrt{|\mu|}$) in the mean 
curves at low frequencies (${\sim}10$~Hz) implying coherent microlensing-induced fluctuations 
in all realizations \citep[rise in \emph{image-2} and dip in \emph{image-3}; see][]{2021MNRAS.508.4869M, 2020PhRvD.101l3512D}.
Since these coherent patterns appear near the lower limit of the LIGO frequency band where 
the detector tends to have decreased sensitivity, it is hard to detect any systematic features 
in the amplification curves.
Nevertheless, with increased sensitivity in the future detectors such as the CE and ET, the 
probability of identifying such microlensing-induced features, at these frequencies, will increase 
which may further help us in identifying the lensed nature of the GW event.
In addition, we find that as we increase the macro-magnification of the image (from bottom-to-top), 
the ${\pm}3\sigma$ region becomes relatively wider implying increased amplitude of microlensing fluctuations.

Another interesting observation is that the scatter is affected more by macro-magnification 
than the underlying microlens density on average.
The \emph{image-1}, with the lowest magnifications, are found in microlens density environments 
spanning nearly an order of magnitude (see Figure~\ref{fig:sl_systems}).
And yet, the mean and the scatter in the amplification curves for those \emph{image-1} systems, 
with very different microlens densities, look barely different from each other (see the bottom-left 
panel of Figure~\ref{fig:ml_system_minima}). 
Any impact on the amplification curves (due to stellar mass microlenses) only 
become prominent in the presence of high macro-magnification.
We note that aforementioned inferences are generally applicable to any minimum or a saddle point. 
In other words, the trends are similar for all of the images, namely, \emph{image-1,2,3} and \emph{4}.

To quantify and better depict the dependence of the scatter on the GW frequency and the 
macro-magnification, in Figure~\ref{fig:ml_linear}, we plot the $3\sigma$ values as a function 
of strong lensing amplification factor ($|\sqrt \mu|$) for all minima (\emph{image-1, 2}) and 
saddle points (\emph{image-3, 4}) in the left and right panels, respectively. 
As before, the trends are similar for both minima and saddle points, that is, the scatter roughly 
increases linearly with macro-magnification whereas the slope of this linear curve becomes steeper 
with increasing frequency. 

Although in this work we apply a detection threshold on the second brightest image, our 
results are valid for any SNR threshold as long as a given lensed image satisfies the criteria 
applied on the macro-magnification and the microlens density, i.e.,
$\{|\mu|, \Sigma_\bullet\} {\leq} \{10, 10^3~{\rm M}_\odot/{\rm pc}^2\}$.
It is because lowering (raising) the SNR threshold will increase (decrease) the number of detectable 
lens systems but the corresponding lensed images will still occupy the same region in the 
$|\mu|-\Sigma_\bullet$ plane as shown in Figure~\ref{fig:sl_systems}.

\section{(Mis)match Analysis}
\label{sec:mismatch}

In this section, we study the magnitude of the microlensing effects on GW signals by 
calculating the (mis)match between lensed and unlensed GW signal. 
First, we generate unlensed waveforms using the \textsc{PyCBC} \citep{2016CQGra..33u5004U} 
with \textsc{IMRPhenomPv2} approximant and with  a lower frequency cut-off ($f_{\rm low}$) 
at 20~Hz. 
We further assume that both BBH components are spin-less and the orbit is circular.
Next, we produce the lensed waveforms by modifying the unlensed waveforms in the 
frequency-domain according to a given $F(f)$.

We then calculate the (mis)match as follows.
If in the absence of lensing, the GW signal is given by $h_{\rm U} (f)$ in frequency domain 
then the strongly lensed GW signal will be $h_{\rm SL} (f) = \sqrt{|\mu|} h_{\rm U} (f)$ and 
strong + microlensed signal will be $h_{\rm ML} (f) = F(f) h_{\rm U} (f)$.
The match ($\mathcal{M}$) between lensed, $h_{\rm L}$ (which can be either be $h_{\rm SL}$ or 
$h_{\rm ML}$), and unlensed waveform ($h_{\rm U}$) is given as 
\citep[e.g.,][]{2016CQGra..33u5004U}
\begin{equation}
    \mathcal{M} = \mathop{max}_{t_0, \phi_0} \frac{\langle h_{\rm L} | h_{\rm U} \rangle}
    {\sqrt{\langle h_{\rm L} | h_{\rm L} \rangle \langle h_{\rm U} | h_{\rm U} \rangle}},
    \label{eq:match}
\end{equation}
where $t_0$ and $\phi_0$ are arrival time and phase of $h_{\rm UL}$, respectively.
The inner product is noise-weighted and defined as
\begin{equation}
    \langle h_1 | h_2 \rangle = 4{\rm Re} \int_{f_{\rm low}}^{f_{\rm {high}}} df 
    \frac{h_1^*(f) h_2(f)}{S_n(f)},
    \label{eq:inner}
\end{equation}
where $S_n(f)$ is the single-sided power spectral density of the detector noise.
From Equation~\ref{eq:match}, we can see that the mismatch between unlensed ($h_{\rm U}$) 
and strongly lensed ($h_{\rm SL}$) waveform will be zero as strong lensing only 
(de)amplifies the unlensed waveform by a constant factor, $\sqrt{|\mu|}$.
Hence, the mismatch between $h_{\rm U}$ and $h_{\rm ML}$ will only arise due to features 
introduced by microlensing.
A high match ($\mathcal{M}>99\%$) generally implies that any deviations due to microlensing 
are weak and will go undistinguished~\citep[e.g.,][]{2014PhRvD..90f2003C}.  
We note that the above mentioned match is between lensed and unlensed waveform and not
the maximum match ($\mathcal{M}_{\rm max}$) which represents the maximum match against a 
template bank.
In general, the $\mathcal{M}_{\rm max}$ is expected to be larger than $\mathcal{M}$.

%%%%%%%%%%%%%%%%%%%%%%%%%%%%%%%%%%%%%%%%%%%%%%%%%%%%%%%%%%%%%%%%%%%%%%%%%%%%%%%%%%%%%%%%%%%
\begin{figure*}
    \centering
    \includegraphics[width=6.3cm, height=4cm]{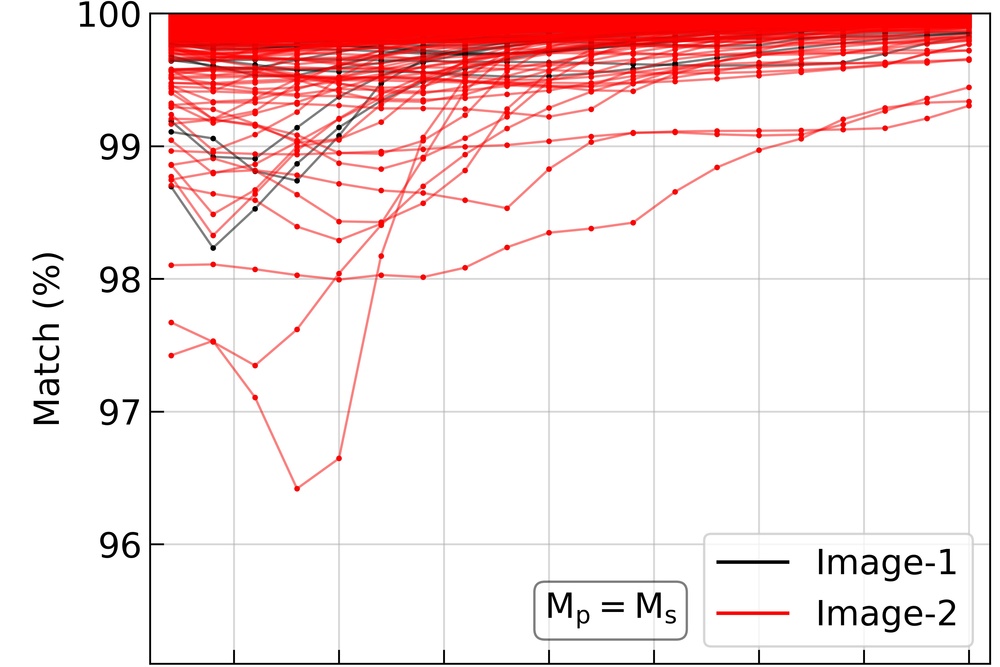}
    \includegraphics[width=5.2cm, height=4cm]{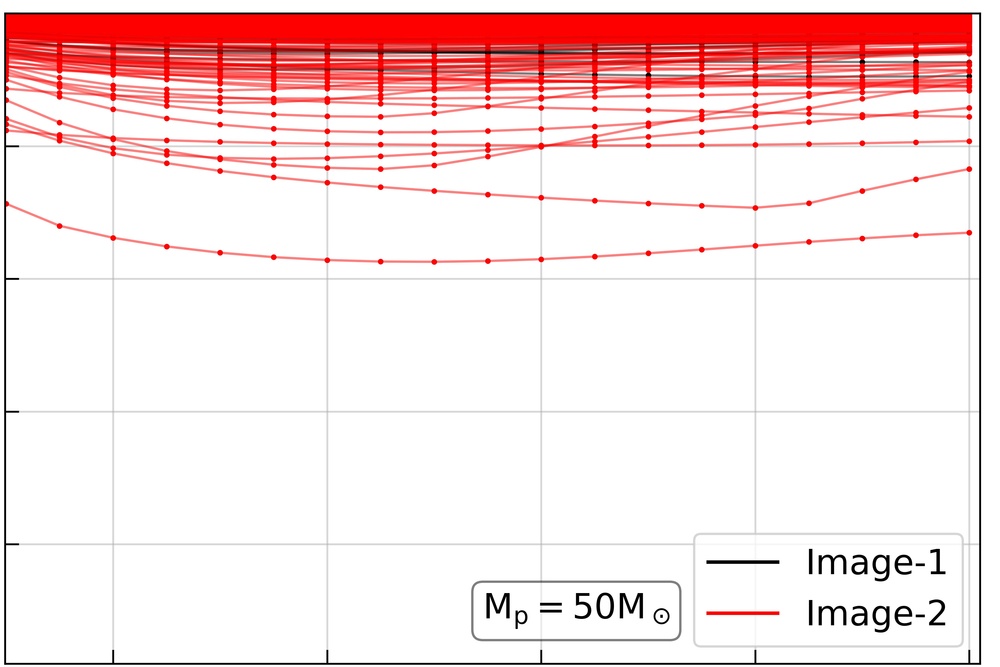}
    \includegraphics[width=5.2cm, height=4cm]{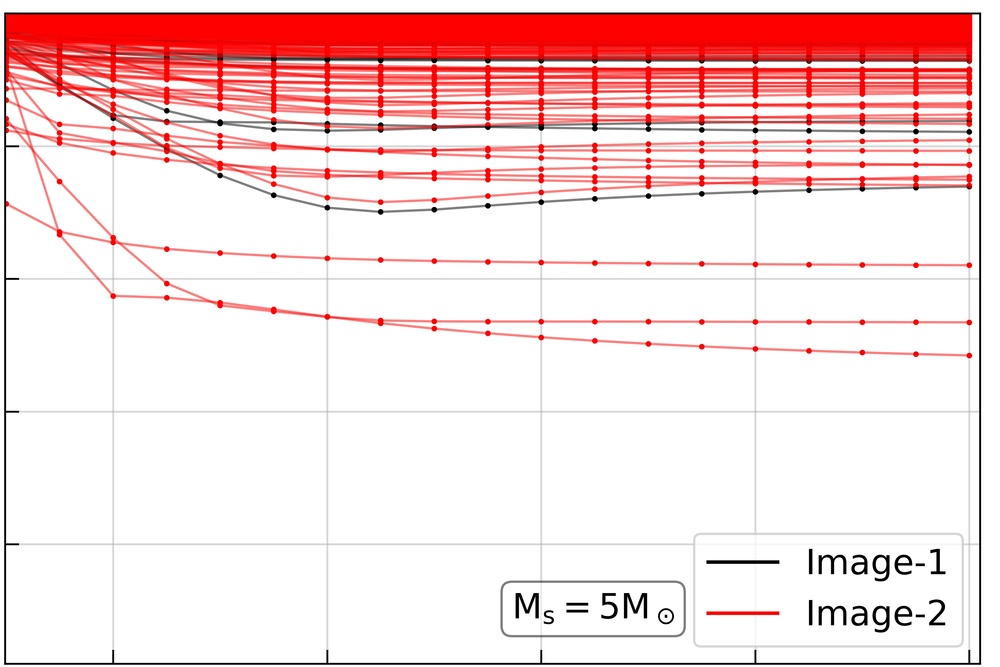}
    
    \includegraphics[width=6.3cm, height=4cm]{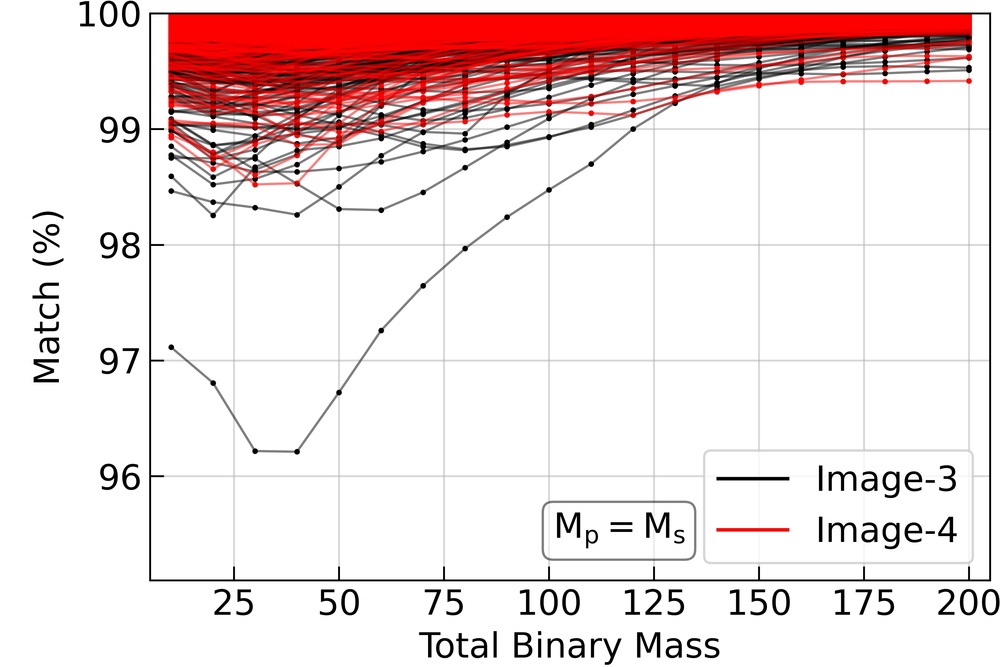}
    \includegraphics[width=5.2cm, height=4cm]{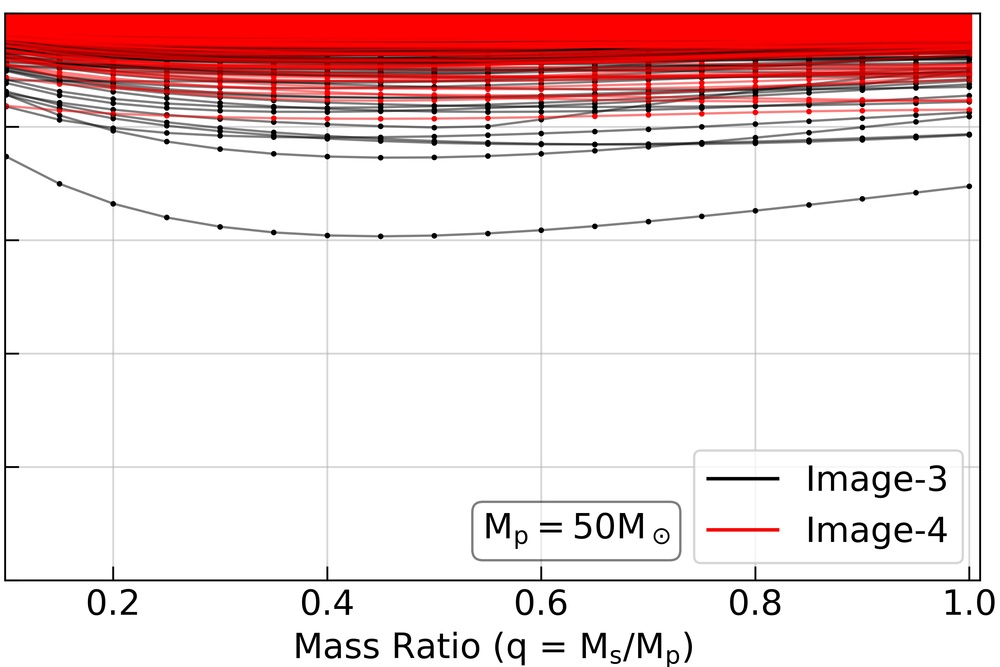}
    \includegraphics[width=5.2cm, height=4cm]{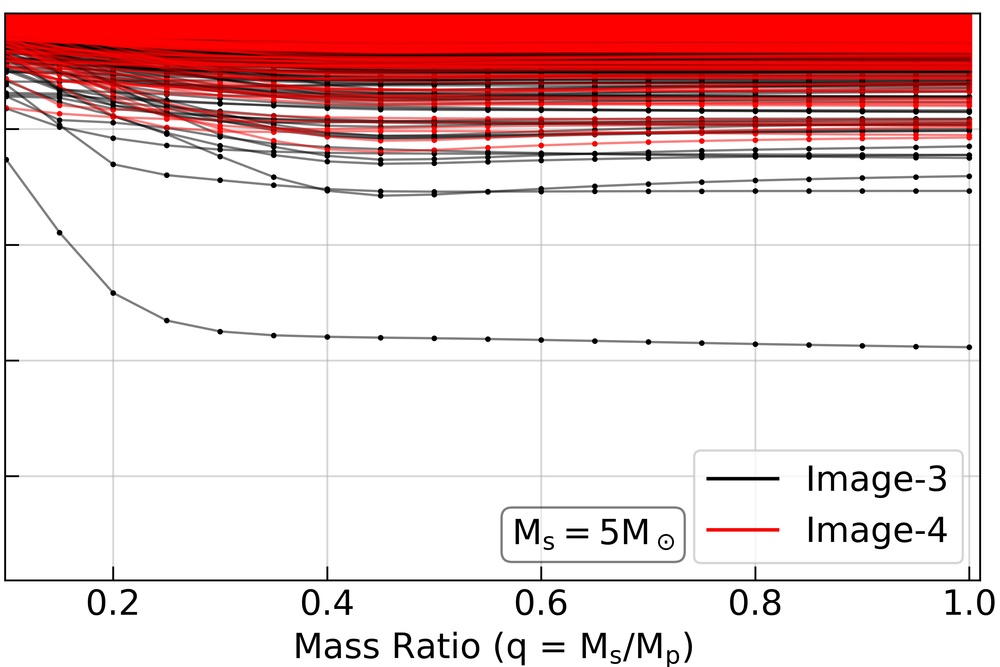}
    
    \caption{Match between unlensed and lensed GW signals as a function of the BBH total mass 
    (left column) and the mass ratio (middle and right columns). The upper and lower rows show 
    results for \emph{image-1, 2} and \emph{image-3, 4}, respectively. For each image, results 
    are shown for 2500 microlensed signals (50 quads $\times$ 50 realizations). Assuming equal 
    masses for the primary ($M_{\rm p}$) and the secondary ($M_{\rm s}$) components of a BBH 
    (left column), the mismatch is found to be worse for low-mass BBHs compared to higher mass BBHs.  
    For a fixed mass of $M_{\rm p}=50~{\rm M}_\odot$ for the primary (middle column), we find 
    that the match does not vary significantly as a function of the mass ratio of the BBH 
    although it is marginally better for systems with extreme mass ratios. Similar trends are 
    seen when varying the mass ratio while the secondary is fixed to $M_{\rm s}=5~{\rm M}_\odot$ 
    (right column).}
    \label{fig:match}
\end{figure*}
%%%%%%%%%%%%%%%%%%%%%%%%%%%%%%%%%%%%%%%%%%%%%%%%%%%%%%%%%%%%%%%%%%%%%%%%%%%%%%%%%%%%%%%%%%%

In Figure~\ref{fig:match}, we show the match for all of the images from a quad (top row 
for \emph{image-1, 2} and bottom row for \emph{3, 4}). 
As we have 50 quad systems and for each image, we simulate 50 realizations, each panel 
contains a total of 2500 curves corresponding to each image.
In the left panel, we vary the BBH total masses assuming both components have equal mass, 
i.e., $M_{\rm p} = M_{\rm s}$, where $M_{\rm p}$ and $M_{\rm s}$ are primary and secondary 
component masses. 
In the middle (right) panel, we vary the mass ratios while fixing the primary (secondary) 
component mass to $50\:(5)\:{\rm M}_\odot$. 
The general conclusion from Figure~\ref{fig:match} is that the match is always greater 
than 99\% for most of the realizations and only a handful (${<}1\%$) of realizations shows 
match less than 99\%. 
This implies that the GW detection pipelines are unlikely to miss any GW signals that are 
affected by microlensing.
Hence, we do not expect any significant bias in the estimated values of the BBH parameters.
We note that $\mathcal{M}_{\rm max}<97\%$ is a detection threshold typically used by GW 
detection pipelines rather than a threshold of $\mathcal{M}<99\%$ and, as mentioned above,
$\mathcal{M}_{\rm max}$ is expected to be larger than $\mathcal{M}$.
Hence, $\mathcal{M}~{\sim}~99\%$ is anticipated to lead $\mathcal{M}_{\rm max}\gtrsim99\%$.

Additionally, in Figure~\ref{fig:match}, the match value increases as we increase the mass 
of the binary components.
This is because massive binaries mainly emit GWs at low frequencies and the amplitude of 
the microlensing effects is weak at such low frequencies although it gets stronger with 
increasing frequency. 
Apart from this, in general, we also observe that \emph{image-1} shows a higher match 
compared to other images.
This can be understood from the fact \emph{image-1} is the global minimum with 
low-to-moderate macro-magnification and it forms in relatively low microlens density regions 
compared to other images. 
On the other hand, \emph{image-4}, which is closest to the lens center, in general, gets 
de-magnified leading to match values similar to \emph{image-1}.
As expected, \emph{image-2} and \emph{image-3} tend to have higher mis-match 
(i.e. $\mathcal{M} <99\%$) more often than \emph{image-1} and \emph{image-4}.

The above observations imply that if the macro-magnification is~$\leq10$ then 
the probability of mismatch between lensed and unlensed signal being worse 
than 1\% is less than 0.01. 
This probability further decreases for \emph{image-1} and (nearly) always remains 
unwavered across multiple realisations of the microlens population.  
Hence, as a first order approximation, we can safely assume negligible microlensing 
effects from stellar mass microlenses in strongly lensed GW signals with 
$|\mu|\leq10$, especially, for \emph{image-1}.
The advantage here is that, among a pair or triplet, identified as a candidate lens 
system, if we can detect the \emph{image-1} then it can be used as a reference GW signal 
to estimate the microlensing effects in other lensed counterparts. 
In \citet{2021arXiv211003308S}, a similar idea is explored to show improvement in the 
parameter estimation of a single microlens. 
However, to study the microlensing effects due to a microlens population in a strongly 
lensed GW candidate, creating many realizations of the microlens population will
probably become imperative.

Although less probable, in rare cases, microlensing due to stellar mass objects
can lead to notable effects even in the~$|\mu|\leq10$ regime (see
Figure~\ref{fig:match}). 
This can happen (not always though) if one or a few microlenses lie around an 
Einstein-radius distance from the patch center in the lens plane.
We find that, all realizations leading to low match (${<}98\%$) in Figure~\ref{fig:match}, 
have a massive microlens with mass ${\sim}20\:{\rm M}_\odot$ located around  1--5 Einstein 
radius (of the microlens) away from the patch center.
Hence, our results imply that, in general, microlensing due to stellar-mass
microlenses at low macro-magnification leads to negligible mismatch
\citep[e.g.,][]{2021MNRAS.503.3326C}, however, presence of microlenses near the patch 
center may lead to significant mismatch even if the macro-magnification is 
small.

\begin{figure*}
    \centering
    \includegraphics[width=8.0cm, height=6.5cm]{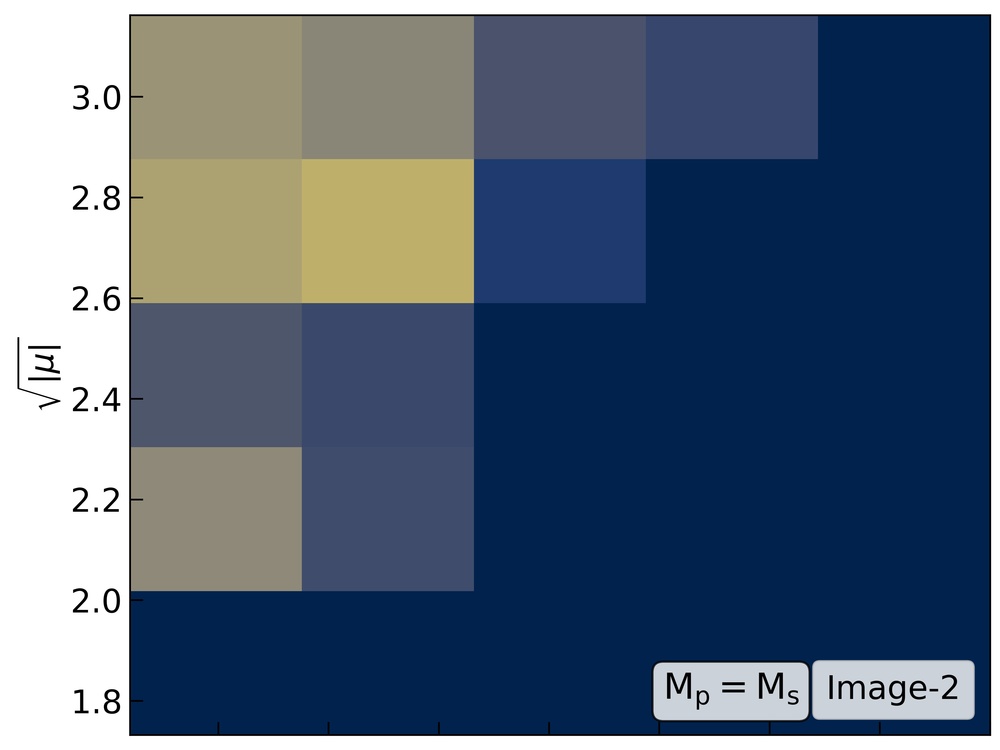}
    \includegraphics[width=8.5cm, height=6.5cm]{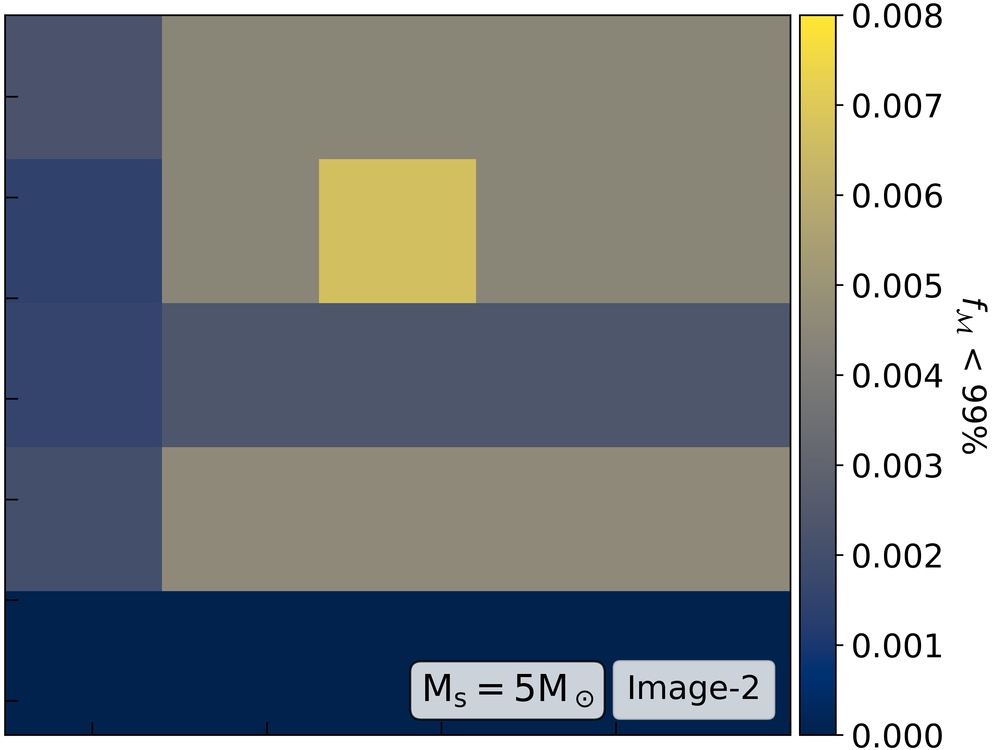}
    \includegraphics[width=8.0cm, height=7.0cm]{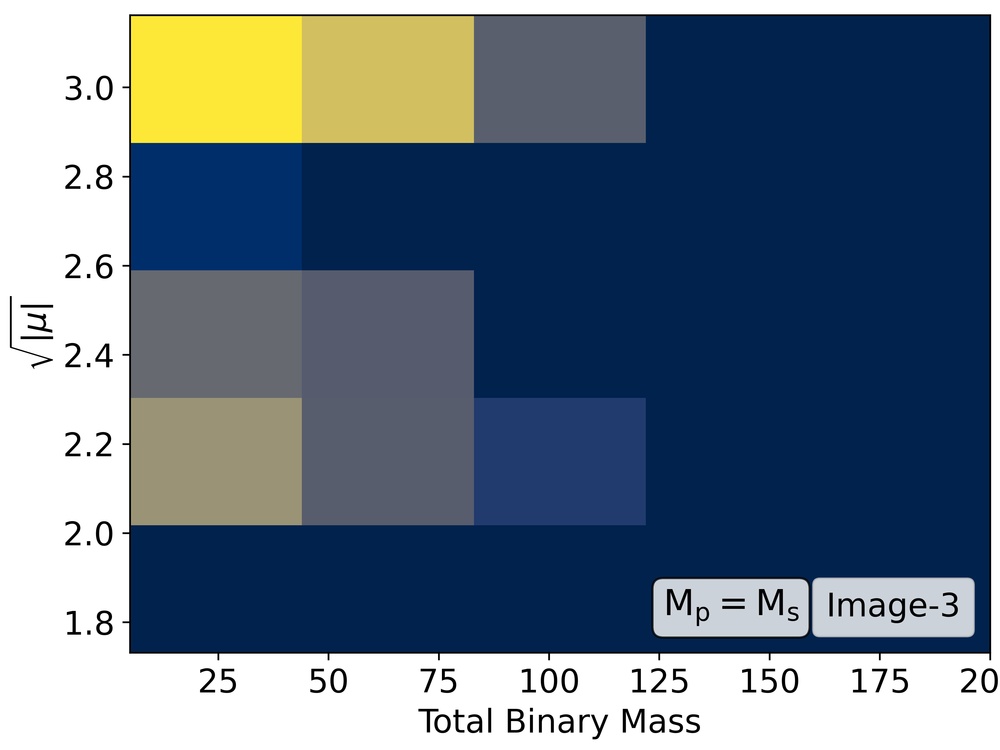}
    \includegraphics[width=8.5cm, height=7.0cm]{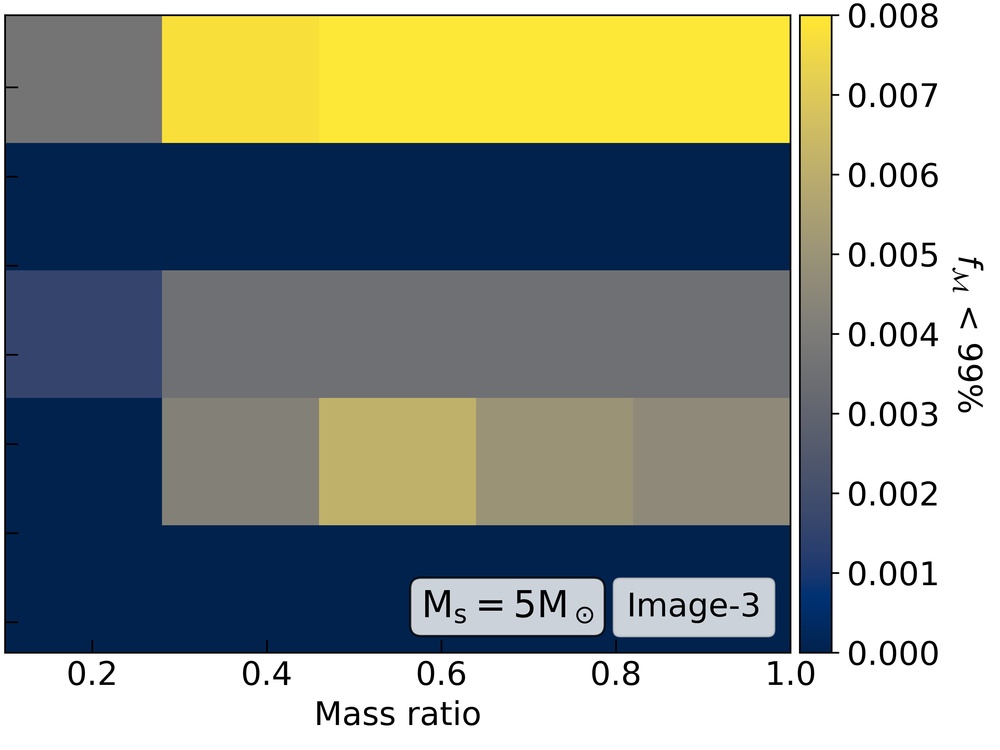}
    \caption{Fraction of images with match~${<}99\%$  ($f_{\mathcal{M}}$)
    as a function of the macro-magnification (y-axes) and the total binary mass (x-axis) 
    for the column on the left and as a function of the mass ratio (x-axis) for the column on 
    the right. The top and bottom rows are for \emph{image-2} and \emph{image-3}, respectively. 
    Mismatch is higher for higher macro-magnifications and for low-mass binaries.}
    \label{fig:pixel_match}
\end{figure*}

We determine the probability of seeing a match below the threshold of 99\% as a function 
of the macro-magnification and the BBH properties, for example, the total BBH mass and the 
mass ratio. 
We calculate these probabilities for all \emph{image-1,2,3,4} but choose to show the results 
only for \emph{image-2} and \emph{image-3} in Figure~\ref{fig:pixel_match} where the mismatch is 
a bit more significant than \emph{image-1} and \emph{image-4}. 
We find that larger fractions of images show a high mismatch at low-mass BBHs (left column), 
particularly, at high macro-magnifications. 
We do not see significant dependency on the BBH mass ratio for the fraction of images (see right 
column and Figure~\ref{fig:match}). 
We find that macro-magnification plays the most important role. 
The trends are similar in both \emph{image-2} and \emph{image-3}.

%%%%%%%%%%%%%%%%%%%%%%%%%%%%%%%%%%%%%%%%%%%%%%%%%%%%%%%%%%%%%%%%%%%%%%%%%%%%%%%%%%%%%%%%%%%
%%%%%%%%%%%%%%%%%%%%%%%%%%%%%%%%%%%%%%%%%%%%%%%%%%%%%%%%%%%%%%%%%%%%%%%%%%%%%%%%%%%%%%%%%%%

\section{High Microlens Density}
\label{sec:high_stelden}

So far our analysis was focused on strongly lensed systems with 
$\{|\mu|, \Sigma_\bullet\}\leq \{10,~10^3~{\rm M}_\odot/{\rm pc}^2\}$.
However, we can see from Figure~\ref{fig:sl_systems} that this does not cover the allowed
$|\mu|-\Sigma_\bullet$ parameter space completely. 
The \emph{image-2} and \emph{image-3} image-pair can attain very high macro-magnifications
($|\mu|{>}100$) for a BBH source located near the caustic in the source plane due to its 
point source nature. 
On the other hand, \emph{image-4} can form in the lens plane where the microlens densities 
can go above $10^3~{\rm M}_\odot/{\rm pc}^2$. 
In this section, we study the microlensing effects in \emph{image-4} with 
$\Sigma_\bullet > 10^3~{\rm M}_\odot/{\rm pc}^2$.
We select four individual \emph{image-4} images (see maroon stars in 
Figure~\ref{fig:sl_systems})  with microlens densities in the range 
$(10^3, 10^4)~{\rm M}_\odot/{\rm pc}^2$. 
The corresponding macro-magnifications ($|\mu|$) lie in the range (0.1, 0.6). 
For each of the four images, we again simulate 50 realizations with a patch-size of 2~pc~$\times$~2~pc. 

\begin{figure}
    \centering
    \includegraphics[height=8.5cm, width=8.5cm]{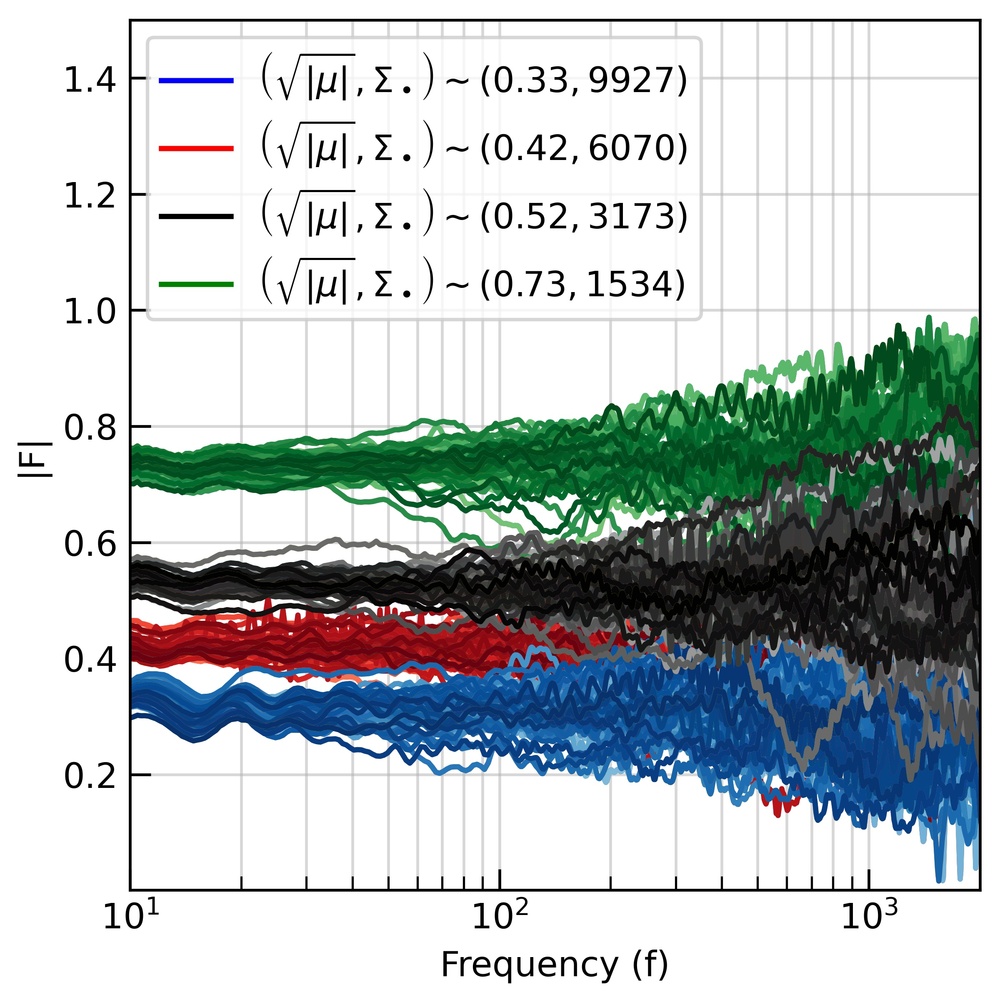}
    \caption{Microlensing effects seen in the saddle points of \emph{image-4} in high density 
    regions. The amplification factor ($|F|$) is shown as a function of frequency for four  
    different combinations of $\{|\mu|,\:\Sigma_\bullet\}$ (corresponding to maroon stars in
    Figure~\ref{fig:sl_systems}). There are 50 realizations for each of the four cases 
    and their respective macro-magnifications and microlens densities are given in the legend.  
    The amplitude of the distortions due to microlensing is negligible despite the high 
    microlens densities owing to the low macro-magnification of these images. 
    See Section~\ref{sec:high_stelden} for more details.}
    \label{fig:high_density}
\end{figure}

The resulting absolute amplification ($|F|$) curves as a function of GW 
frequencies are shown in Figure~\ref{fig:high_density}. 
For all four cases, we see two different kind of features: 
(i) random oscillations unique to each realization with increasing amplitude as we go 
from low to high frequencies. 
(ii) coherent oscillations across all realizations of image with 
$\Sigma_\bullet {=} 9927~{\rm M}_\odot/{\rm pc}^2$ at low frequencies (${\sim}20$ Hz).
The random oscillations are genuine features arising due to the presence of microlens 
population and unique in each realizations as the microlenses are randomly distributed.
The coherent oscillations, at low frequencies, are artifacts and are arising due to the 
finite patch size. 
It turns out that a small patch can lead to the formation of spurious micro-images at 
large time delays which leads to frequency-dependent oscillations at low frequencies. 
If we simulate a bigger patch, these oscillations will disappear. 
But, simulating a large patch implies an increase in the number of microlenses leading 
to increase in the computational times. 
Apart from that, the $|F|$ curves only show variation of $\sim 0.2-0.3$ over the mean
magnification which tends to be less than one and thus, the amplitude of these 
oscillations is not significant across LIGO--Virgo frequencies (10~Hz to $10^3$~Hz) 
resulting in negligible effects in the mismatch as well as the parameter estimation. 
Considering these factors, we choose not to simulate bigger patches in the lens plane 
as it is expected to not affect our conclusion. 
We conclude that the microlensing-driven frequency-dependent effects (due to stellar 
mass population) are negligible in de-magnified lensed GW signals located in high 
microlens density regions within the lens galaxy.

\section{Two image systems (Doubles)}
\label{sec:two_image}

\begin{figure}
    \centering
    \includegraphics[height=8.5cm, width=8.5cm]{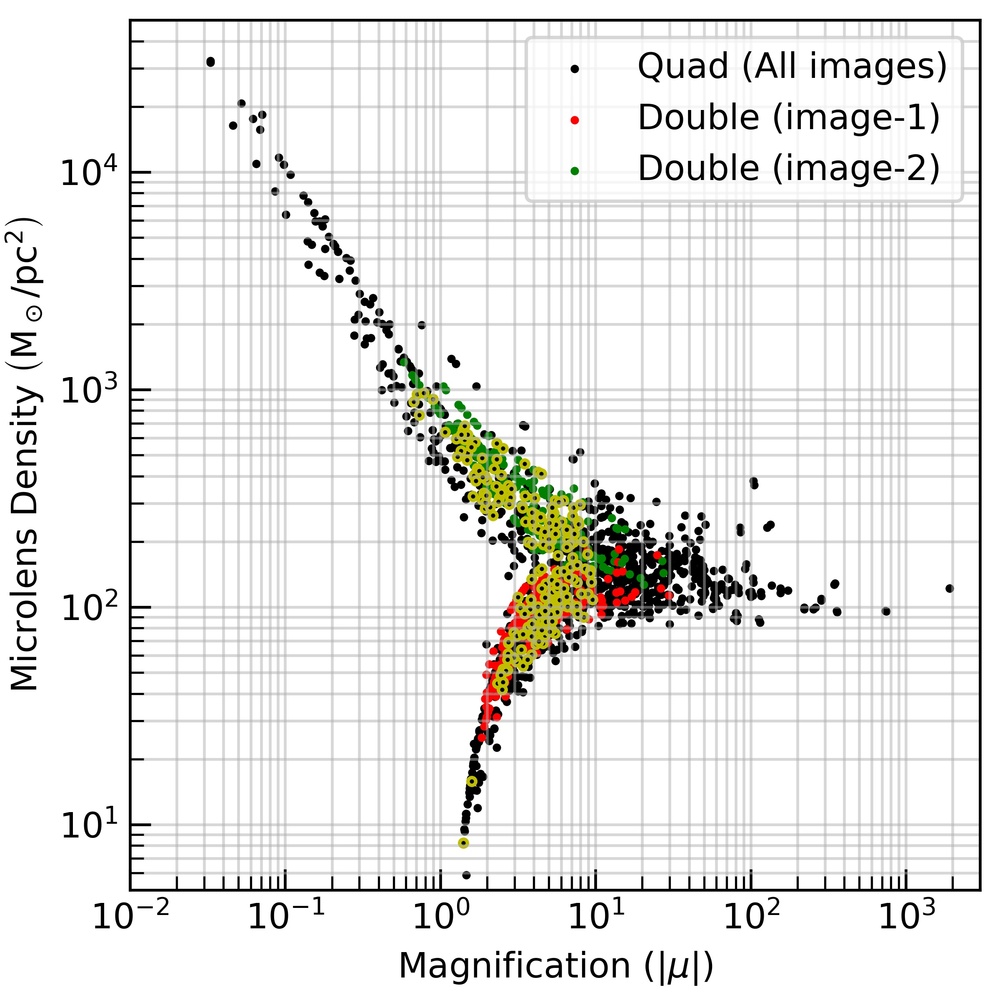}
    \caption{Distributions of doubly lensed images in the plane of macro-magnification~($|\mu|$)
    and microlens density~($\Sigma_\bullet$) as compared to the distributions seen in the lensed 
    images of quads. The black dots represent all four images (\emph{image-1, 2, 3, 4}) of 
    the simulated quads (same as the quads in Figure~\ref{fig:sl_systems}). The red and green 
    points represent the minima (\emph{image-1}) and saddle points (\emph{image-2}) belonging 
    to doubles, respectively. The yellow-edge points denote the 50 quads used in our analysis. 
    The overlap between quads and doubles in this plane implies the minima and saddle points of 
    doubles have similar local environments and thus, results obtained for quads hold true for doubles.}
    \label{fig:double_quad}
\end{figure}

Until now, we have been investigating quads but since the SIE lens models produce both 
doubles and quads which is also consistent with real observations of lens samples, we 
briefly discuss implications of our results on images belonging to doubles.
Fortunately, the \emph{image-1, 2} of doubles have a significant overlap with the images 
in quads in the $|\mu|-\Sigma_\bullet$ plane. 
We verify this by over-plotting 200 doubles generated in a similar way to the quads in the
$|\mu|-\Sigma_\bullet$ plane as shown in Figure~\ref{fig:double_quad}. 
Note that the underlying quad population is the same as shown in 
Figure~\ref{fig:sl_systems} except that all four images (\emph{image-1, 2, 3, 4}) from 
a quad are shown by a single black color for reference and as before, the yellow edge 
circles highlight the 50 systems used in our analysis. 
The red and green points now show the \emph{image-1, 2} corresponding to minima and 
saddle points found in doubles.

This extensive overlap implies that the minima and saddle points formed in the doubles 
and quads are found in similar lens environments, that is, similar combinations of 
macro-magnifications and microlens densities leading to a similar \emph{butterfly}-like 
distribution in the $|\mu|$--$\Sigma_\bullet$ plane. 
It is then safe to use the minima and saddle point realizations from the sample of 
quads to deduce the microlensing effects expected to be seen in a sample of doubles.
As a result, our conclusions inferred from the images belonging to quads are also 
applicable to images belonging to doubles. 

This is an important result to be noted because from an observational point of view, 
there tends to be an ambiguity whether a pair of candidate lensed events belong to a 
double or a quad. 
Only when the image types of the pair are robustly known and are expected to have the 
same type (i.e., they both are either minima or saddle points) can we be certain that 
the pair of images are from a quad. 
However, since the degree of microlensing effects are similar for both doubles and 
quads, it is not required to know beforehand the image types or the true multiplicity 
of a lens system (i.e. a double or a quad). 

We see that in our sample of doubles, only a couple of lens systems have saddle points 
with microlens densities ${>}10^3~{\rm M}_\odot$.
This is because increasing density will decrease the strong lens magnification which 
leads the SNR to drop below 8. 
However, as discussed in Section~\ref{sec:high_stelden}, the amplitude of microlensing 
effects further decreases.
This only depends on the strong lensing information, hence, the results from 
Section~\ref{sec:high_stelden} can again be extrapolated for double image systems.

%%%%%%%%%%%%%%%%%%%%%%%%%%%%%%%%%%%%%%%%%%%%%%%%%%%%%%%%%%%%%%%%%%%%%%%%%%%%%%%%%%%%%%%%%%%
%%%%%%%%%%%%%%%%%%%%%%%%%%%%%%%%%%%%%%%%%%%%%%%%%%%%%%%%%%%%%%%%%%%%%%%%%%%%%%%%%%%%%%%%%%%

%%%%%%%%%%%%%%%%%%%%%%%%%%%%%%%%%%%%%%%%%%%%%%%%%%%%%%%%%%%%%%%%%%%%%%%%%%%%%%%%%%%%%%%%%%%
%%%%%%%%%%%%%%%%%%%%%%%%%%%%%%%%%%%%%%%%%%%%%%%%%%%%%%%%%%%%%%%%%%%%%%%%%%%%%%%%%%%%%%%%%%%

\section{Discussion and Conclusions}
\label{sec:conclusions}

The goal of this work is to study realistic samples of galaxy-scale lens systems in order 
to assess the probability of microlensing arising due to realistic stellar populations 
and to understand the extent to which microlensing could affect the detectability of 
strongly lensed GW events. 
We investigated the effect of microlensing on two types of lensed images, minima and 
saddle points, found in typical galaxy-scale lenses where the images have low-to-moderate
macro-magnifications ($|\mu|=(0.5, 10)$) and are located in typical microlens density 
environments $(\Sigma_\bullet = (10, 10^3)~{\rm M}_\odot/{\rm pc}^2)$ in the LIGO frequency 
band ($10$~Hz -- $10^4$~Hz). 

In our earlier work \citet{2021MNRAS.508.4869M}, we studied the microlensing effects for individual
minima and saddle points considering specific macro-magnification and microlens density 
values appropriate for galaxy lenses, however, the underlying goal was to identify which 
parameters influence microlensing the most and how severe the microlensing effects could 
be for those specific scenarios. 
In the current work, we are studying the lens population as a whole to determine the 
fraction of galaxy-scale lenses that could be affected by microlensing and to what extent.  
To do so, we considered a total of 50 quadruply-imaged lens systems (quads) where the 
second brightest image has a three-detector network-SNR greater than 8 (corresponding to 
the LIGO-O3 sensitivity).
For each strongly lensed image within a quad, we generated 50 realizations to study the 
scatter (introduced due to microlensing) in the amplification factor. 
We also calculated the mismatch between the lensed and unlensed GW signals to understand 
how microlensing will distort and modulate the individual lensed signal and different 
lensed GW counterparts of a signal.
Furthermore, the \emph{image-4}, which forms near the lens center, tends to be located 
in high microlens density regions ($\Sigma_\bullet{>}10^3~{\rm M}_\odot/{\rm pc}^2$). 
Since understanding the microlensing effects in such extreme conditions is also important 
but computationally expensive, we considered only four high microlens density cases by 
generating 50 realizations for each case, as done earlier.
In addition to the quads, we also expect to find doubly imaged lenses (doubles) among 
the galaxy-scale lens systems. 
Even though we only studied images belonging to quads in this work, we discussed their 
similarity with the images in doubles and validity of our results for doubles.

The main conclusions of our work are described below.
\begin{itemize}
    \item  We find that the microlensing effects are more sensitive to the 
    macro-magnification than the underlying microlens density (constituting stellar mass 
    objects) similar to \citet{2021MNRAS.508.4869M}. Even when the images are located 
    in extremely high microlens density regions~($>10^3~{\rm M}_\odot/{\rm pc}^2$), the 
    microlensing distortions are minimal due to low macro-magnification.
    
    \item  For both minima and saddle points with macro-magnification~${\leq}~10$, the
    microlensing due to stellar-mass objects does introduce frequency-dependent effects. 
    However, in ${>}~99\%$ cases, the mismatch (between lensed and unlensed GW signals) 
    remains ${<}~1\%$ as  the amplitude of microlensing distortions is not significant 
    enough. Thus, the detectability of any of the strongly lensed images with~$|\mu|\leq10$ 
    is unlikely to be affected by microlensing due to stellar mass objects~\citep{2021MNRAS.503.3326C}. 
    This also implies that at $|\mu|\leq10$ we expect negligible bias due to microlensing 
    in parameter estimation and identification of different lensed 
    counterparts alleviating the concern highlighted in~\citet{2020MNRAS.492.1127M}.
    
    \item The strongly lensed GW signal corresponding to global minima (\emph{image-1}) 
    is the least affected by microlensing for all total masses and mass ratios of the BBH. 
    Hence, if any of the remaining lensed images are identified together with \emph{image-1}, 
    we can use the \emph{image-1} as an uncontaminated ``reference" image to estimate
    unambiguously the microlensing-specific effects on the other lensed images and hence, 
    potentially, their microlensing model parameters too.

    \item For a given strongly lensed signal, the microlensing distortions increase with
    increasing the GW frequency (as also noted in \citealt{2021MNRAS.508.4869M}). Hence, 
    a low-mass BBH, wherein the binary coalesce at higher frequencies, is expected to show 
    more significant microlensing effects compared to a high-mass BBH. 

    \item In general, since the \emph{image-4} is de-magnified, it will tend to be below 
    the threshold of standard GW detection pipelines (i.e. a sub-threshold event). A
    magnification boost from microlensing could make such events to become super-threshold, 
    in principle. However, this is almost impossible in reality (considering the stellar mass 
    population) as the amplitude of the microlensing distortions are weak for all of the 
    $|\mu|\leq10$ \emph{image-4} explored in our analysis. Further de-magnification from 
    microlensing will likely make it even more difficult to find \emph{image-4}, generally.
    
    \item Our conclusions are equally valid for doubles, since their minima and saddle 
    points occupy the same region in the macro-magnification vs microlens density 
    ($|\mu|-\Sigma_\bullet$) plane, as long as the $|\mu|\leq10$.
    
    \item We note that ${\sim}40\%$ of detectable event pairs from quads and and ${\sim}90\%$ 
    of the doubles from our realistic mocks,  corresponding to current LIGO--Virgo sensitivities,  
    meet the $|\mu|\leq10$ cut. For the fourth observing run of LIGO--Virgo, the fractions 
    become~${\sim}50\%$ for quads and remain nearly the same for the doubles. But with future 
    detectors such as the Cosmic Explorer or Einstein Telescope,  these fractions rise up 
    to~${\sim}90\%$ and~${\sim}99\%$ for quads and doubles, respectively.

\end{itemize}

Our conclusions have implications in the detection of strongly lensed GW event pairs. 
Searching methods such as ranking statistic to identify strongly lensed candidate pairs, 
for example, ${\cal M}^{\rm gal}$ (MM21) are based on joint distributions of magnifications 
and time-delays. 
They rely on the fact that there are correlations between the relative magnifications 
and time-delays of a pair of strongly lensed signals. 
If microlensing causes significant distortions to magnifications of one or both of the 
lensed signals, then the ranking statistic based on joint distributions could fail to 
identify a strongly lensed event pair correctly. 
However, our analysis suggests that the relative magnifications will  be unaffected by 
microlensing for almost all lensed signals with $|\mu|\leq10$ and thus, offering a fast 
and robust means of identifying candidate lensed event pairs.

As mentioned earlier, the global minimum (\emph{image-1}) for $|\mu|{\leq}10$ does not get
significantly affected by the microlensing  due to stellar mass microlenses. 
However, high-mass microlenses~(${>}100~{\rm M}_\odot$) may still lead to detectable
frequency-dependent modulations.
The detection of such high-mass BBH sources in LIGO \citep{2021arXiv211103606T} implies 
that such objects are present in galaxies albeit in small numbers probably.  
Hence, \emph{image-1}, if affected by high-mass microlenses, may become an excellent 
probe to study this microlens population in lens galaxies (as the stellar mass microlenses
lead to negligible microlensing-effects) and may allow us to put constraint on their properties.
We plan to investigation this in our future work.

It is very likely that our results will not be applicable for lensed signals with 
macro-magnification~($|\mu|$)~${>}~10$. 
In general, we expect that \emph{image-2} and \emph{image-3} pair to attain very high 
magnification whereas the corresponding \emph{image-1} and \emph{image-4} to have 
low-to-moderate macro-magnification. 
In such cases, the estimated parameters values for \emph{image-2, 3} pair can be 
significantly different than \emph{image-1, 4}~\citep{2020MNRAS.492.1127M}. 
Hence, we will study the microlensing effects in systems where one or more lensed 
counterparts has macro-magnification~$(|\mu|)>10$ in a subsequent paper.

\section{Acknowledgements}
AKM would like to thank the Ministry of Science and Technology, Israel for financial support. 
A. Mishra would like to thank the University Grants Commission (UGC), India, for financial 
support as a research fellow.
Authors gratefully acknowledge the use of high performance computing facilities at IUCAA, Pune.
This research has made use of NASA’s Astrophysics Data System Bibliographic Services.

\section{Data Availability}
The simulated strong lens system data is available from MM21 upon reasonable request.
The microlensing simulation data can be made available upon reasonable request to the 
corresponding author.

%%%%%%%%%%%%%%%%%%%%%%%%%%%%%%%%%%%%%%%%%%%%%%%%%%%%%%%%%%%%
\bibliographystyle{mnras}
\bibliography{reference}

\begin{thebibliography}{}
\makeatletter
\relax
\def\mn@urlcharsother{\let\do\@makeother \do\$\do\&\do\#\do\^\do\_\do\%\do\~}
\def\mn@doi{\begingroup\mn@urlcharsother \@ifnextchar [ {\mn@doi@}
  {\mn@doi@[]}}
\def\mn@doi@[#1]#2{\def\@tempa{#1}\ifx\@tempa\@empty \href
  {http://dx.doi.org/#2} {doi:#2}\else \href {http://dx.doi.org/#2} {#1}\fi
  \endgroup}
\def\mn@eprint#1#2{\mn@eprint@#1:#2::\@nil}
\def\mn@eprint@arXiv#1{\href {http://arxiv.org/abs/#1} {{\tt arXiv:#1}}}
\def\mn@eprint@dblp#1{\href {http://dblp.uni-trier.de/rec/bibtex/#1.xml}
  {dblp:#1}}
\def\mn@eprint@#1:#2:#3:#4\@nil{\def\@tempa {#1}\def\@tempb {#2}\def\@tempc
  {#3}\ifx \@tempc \@empty \let \@tempc \@tempb \let \@tempb \@tempa \fi \ifx
  \@tempb \@empty \def\@tempb {arXiv}\fi \@ifundefined
  {mn@eprint@\@tempb}{\@tempb:\@tempc}{\expandafter \expandafter \csname
  mn@eprint@\@tempb\endcsname \expandafter{\@tempc}}}

\bibitem[\protect\citeauthoryear{{Abbott} et~al.,}{{Abbott}
  et~al.}{2018}]{2018LRR....21....3A}
{Abbott} B.~P.,  et~al., 2018, \mn@doi [Living Reviews in Relativity]
  {10.1007/s41114-018-0012-9}, \href
  {https://ui.adsabs.harvard.edu/abs/2018LRR....21....3A} {21, 3}

\bibitem[\protect\citeauthoryear{{Abbott} et~al.,}{{Abbott}
  et~al.}{2021}]{2021PhRvX..11b1053A}
{Abbott} R.,  et~al., 2021, \mn@doi [Physical Review X]
  {10.1103/PhysRevX.11.021053}, \href
  {https://ui.adsabs.harvard.edu/abs/2021PhRvX..11b1053A} {11, 021053}

\bibitem[\protect\citeauthoryear{{Acernese} et~al.,}{{Acernese}
  et~al.}{2015}]{2015CQGra..32b4001A}
{Acernese} F.,  et~al., 2015, \mn@doi [Classical and Quantum Gravity]
  {10.1088/0264-9381/32/2/024001}, \href
  {https://ui.adsabs.harvard.edu/abs/2015CQGra..32b4001A} {32, 024001}

\bibitem[\protect\citeauthoryear{{Akutsu} et~al.,}{{Akutsu}
  et~al.}{2021}]{2021PTEP.2021eA101A}
{Akutsu} T.,  et~al., 2021, \mn@doi [Progress of Theoretical and Experimental
  Physics] {10.1093/ptep/ptaa125}, \href
  {https://ui.adsabs.harvard.edu/abs/2021PTEP.2021eA101A} {2021, 05A101}

\bibitem[\protect\citeauthoryear{{Bailes} et~al.,}{{Bailes}
  et~al.}{2021}]{2021NatRP...3..344B}
{Bailes} M.,  et~al., 2021, \mn@doi [Nature Reviews Physics]
  {10.1038/s42254-021-00303-8}, \href
  {https://ui.adsabs.harvard.edu/abs/2021NatRP...3..344B} {3, 344}

\bibitem[\protect\citeauthoryear{{Baraldo}, {Hosoya}  \& {Nakamura}}{{Baraldo}
  et~al.}{1999}]{1999PhRvD..59h3001B}
{Baraldo} C.,  {Hosoya} A.,   {Nakamura} T.~T.,  1999, \mn@doi [\prd]
  {10.1103/PhysRevD.59.083001}, \href
  {https://ui.adsabs.harvard.edu/abs/1999PhRvD..59h3001B} {59, 083001}

\bibitem[\protect\citeauthoryear{{Barausse} et~al.,}{{Barausse}
  et~al.}{2020}]{2020GReGr..52...81B}
{Barausse} E.,  et~al., 2020, \mn@doi [General Relativity and Gravitation]
  {10.1007/s10714-020-02691-1}, \href
  {https://ui.adsabs.harvard.edu/abs/2020GReGr..52...81B} {52, 81}

\bibitem[\protect\citeauthoryear{{Belczynski}, {Ryu}, {Perna}, {Berti},
  {Tanaka}  \& {Bulik}}{{Belczynski} et~al.}{2017}]{2017MNRAS.471.4702B}
{Belczynski} K.,  {Ryu} T.,  {Perna} R.,  {Berti} E.,  {Tanaka} T.~L.,
  {Bulik} T.,  2017, \mn@doi [\mnras] {10.1093/mnras/stx1759}, \href
  {https://ui.adsabs.harvard.edu/abs/2017MNRAS.471.4702B} {471, 4702}

\bibitem[\protect\citeauthoryear{{Bird}, {Cholis}, {Mu{\~n}oz},
  {Ali-Ha{\"\i}moud}, {Kamionkowski}, {Kovetz}, {Raccanelli}  \&
  {Riess}}{{Bird} et~al.}{2016}]{2016PhRvL.116t1301B}
{Bird} S.,  {Cholis} I.,  {Mu{\~n}oz} J.~B.,  {Ali-Ha{\"\i}moud} Y.,
  {Kamionkowski} M.,  {Kovetz} E.~D.,  {Raccanelli} A.,   {Riess} A.~G.,  2016,
  \mn@doi [\prl] {10.1103/PhysRevLett.116.201301}, \href
  {https://ui.adsabs.harvard.edu/abs/2016PhRvL.116t1301B} {116, 201301}

\bibitem[\protect\citeauthoryear{{Bontz} \& {Haugan}}{{Bontz} \&
  {Haugan}}{1981}]{1981Ap&SS..78..199B}
{Bontz} R.~J.,  {Haugan} M.~P.,  1981, \mn@doi [\apss] {10.1007/BF00654034},
  \href {https://ui.adsabs.harvard.edu/abs/1981Ap&SS..78..199B} {78, 199}

\bibitem[\protect\citeauthoryear{{Broadhurst}, {Diego}  \&
  {Smoot}}{{Broadhurst} et~al.}{2018}]{2018arXiv180205273B}
{Broadhurst} T.,  {Diego} J.~M.,   {Smoot} George I.,  2018, arXiv e-prints,
  \href {https://ui.adsabs.harvard.edu/abs/2018arXiv180205273B} {p.
  arXiv:1802.05273}

\bibitem[\protect\citeauthoryear{{Bulashenko} \& {Ubach}}{{Bulashenko} \&
  {Ubach}}{2021}]{2021arXiv211210773B}
{Bulashenko} O.,  {Ubach} H.,  2021, arXiv e-prints, \href
  {https://ui.adsabs.harvard.edu/abs/2021arXiv211210773B} {p. arXiv:2112.10773}

\bibitem[\protect\citeauthoryear{{Cao}, {Li}  \& {Wang}}{{Cao}
  et~al.}{2014}]{2014PhRvD..90f2003C}
{Cao} Z.,  {Li} L.-F.,   {Wang} Y.,  2014, \mn@doi [\prd]
  {10.1103/PhysRevD.90.062003}, \href
  {https://ui.adsabs.harvard.edu/abs/2014PhRvD..90f2003C} {90, 062003}

\bibitem[\protect\citeauthoryear{{Cheung}, {Gais}, {Hannuksela}  \&
  {Li}}{{Cheung} et~al.}{2021}]{2021MNRAS.503.3326C}
{Cheung} M. H.~Y.,  {Gais} J.,  {Hannuksela} O.~A.,   {Li} T. G.~F.,  2021,
  \mn@doi [\mnras] {10.1093/mnras/stab579}, \href
  {https://ui.adsabs.harvard.edu/abs/2021MNRAS.503.3326C} {503, 3326}

\bibitem[\protect\citeauthoryear{{Choi}, {Park}  \& {Vogeley}}{{Choi}
  et~al.}{2007}]{2007ApJ...658..884C}
{Choi} Y.-Y.,  {Park} C.,   {Vogeley} M.~S.,  2007, \mn@doi [\apj]
  {10.1086/511060}, \href
  {https://ui.adsabs.harvard.edu/abs/2007ApJ...658..884C} {658, 884}

\bibitem[\protect\citeauthoryear{{Christian}, {Vitale}  \& {Loeb}}{{Christian}
  et~al.}{2018}]{2018PhRvD..98j3022C}
{Christian} P.,  {Vitale} S.,   {Loeb} A.,  2018, \mn@doi [\prd]
  {10.1103/PhysRevD.98.103022}, \href
  {https://ui.adsabs.harvard.edu/abs/2018PhRvD..98j3022C} {98, 103022}

\bibitem[\protect\citeauthoryear{{Dai} \& {Venumadhav}}{{Dai} \&
  {Venumadhav}}{2017}]{2017arXiv170204724D}
{Dai} L.,  {Venumadhav} T.,  2017, arXiv e-prints, \href
  {https://ui.adsabs.harvard.edu/abs/2017arXiv170204724D} {p. arXiv:1702.04724}

\bibitem[\protect\citeauthoryear{{Dai}, {Zackay}, {Venumadhav}, {Roulet}  \&
  {Zaldarriaga}}{{Dai} et~al.}{2020}]{2020arXiv200712709D}
{Dai} L.,  {Zackay} B.,  {Venumadhav} T.,  {Roulet} J.,   {Zaldarriaga} M.,
  2020, arXiv e-prints, \href
  {https://ui.adsabs.harvard.edu/abs/2020arXiv200712709D} {p. arXiv:2007.12709}

\bibitem[\protect\citeauthoryear{{Deguchi} \& {Watson}}{{Deguchi} \&
  {Watson}}{1986}]{1986PhRvD..34.1708D}
{Deguchi} S.,  {Watson} W.~D.,  1986, \mn@doi [\prd]
  {10.1103/PhysRevD.34.1708}, \href
  {https://ui.adsabs.harvard.edu/abs/1986PhRvD..34.1708D} {34, 1708}

\bibitem[\protect\citeauthoryear{{Diego}}{{Diego}}{2019}]{2019A&A...625A..84D}
{Diego} J.~M.,  2019, \mn@doi [\aap] {10.1051/0004-6361/201833670}, \href
  {https://ui.adsabs.harvard.edu/abs/2019A&A...625A..84D} {625, A84}

\bibitem[\protect\citeauthoryear{{Diego}}{{Diego}}{2020}]{2020PhRvD.101l3512D}
{Diego} J.~M.,  2020, \mn@doi [\prd] {10.1103/PhysRevD.101.123512}, \href
  {https://ui.adsabs.harvard.edu/abs/2020PhRvD.101l3512D} {101, 123512}

\bibitem[\protect\citeauthoryear{{Diego}, {Hannuksela}, {Kelly}, {Pagano},
  {Broadhurst}, {Kim}, {Li}  \& {Smoot}}{{Diego}
  et~al.}{2019}]{2019A&A...627A.130D}
{Diego} J.~M.,  {Hannuksela} O.~A.,  {Kelly} P.~L.,  {Pagano} G.,  {Broadhurst}
  T.,  {Kim} K.,  {Li} T.~G.~F.,   {Smoot} G.~F.,  2019, \mn@doi [\aap]
  {10.1051/0004-6361/201935490}, \href
  {https://ui.adsabs.harvard.edu/abs/2019A&A...627A.130D} {627, A130}

\bibitem[\protect\citeauthoryear{{Eldridge}, {Stanway}, {Xiao}, {McClelland},
  {Taylor}, {Ng}, {Greis}  \& {Bray}}{{Eldridge}
  et~al.}{2017}]{2017PASA...34...58E}
{Eldridge} J.~J.,  {Stanway} E.~R.,  {Xiao} L.,  {McClelland} L.~A.~S.,
  {Taylor} G.,  {Ng} M.,  {Greis} S.~M.~L.,   {Bray} J.~C.,  2017, \mn@doi
  [\pasa] {10.1017/pasa.2017.51}, \href
  {https://ui.adsabs.harvard.edu/abs/2017PASA...34...58E} {34, e058}

\bibitem[\protect\citeauthoryear{{Evans} et~al.,}{{Evans}
  et~al.}{2021}]{2021arXiv210909882E}
{Evans} M.,  et~al., 2021, arXiv e-prints, \href
  {https://ui.adsabs.harvard.edu/abs/2021arXiv210909882E} {p. arXiv:2109.09882}

\bibitem[\protect\citeauthoryear{Janquart, Seo, Hannuksela, Li  \&
  Broeck}{Janquart et~al.}{2021}]{Janquart_2021}
Janquart J.,  Seo E.,  Hannuksela O.~A.,  Li T. G.~F.,   Broeck C. V.~D.,
  2021, \mn@doi [The Astrophysical Journal Letters] {10.3847/2041-8213/ac3bcf},
  923, L1

\bibitem[\protect\citeauthoryear{{Kawamura} et~al.,}{{Kawamura}
  et~al.}{2021}]{2021PTEP.2021eA105K}
{Kawamura} S.,  et~al., 2021, \mn@doi [Progress of Theoretical and Experimental
  Physics] {10.1093/ptep/ptab019}, \href
  {https://ui.adsabs.harvard.edu/abs/2021PTEP.2021eA105K} {2021, 05A105}

\bibitem[\protect\citeauthoryear{{Kinugawa}, {Miyamoto}, {Kanda}  \&
  {Nakamura}}{{Kinugawa} et~al.}{2016}]{2016MNRAS.456.1093K}
{Kinugawa} T.,  {Miyamoto} A.,  {Kanda} N.,   {Nakamura} T.,  2016, \mn@doi
  [\mnras] {10.1093/mnras/stv2624}, \href
  {https://ui.adsabs.harvard.edu/abs/2016MNRAS.456.1093K} {456, 1093}

\bibitem[\protect\citeauthoryear{{Kormann}, {Schneider}  \&
  {Bartelmann}}{{Kormann} et~al.}{1994}]{1994A&A...284..285K}
{Kormann} R.,  {Schneider} P.,   {Bartelmann} M.,  1994, \aap, \href
  {https://ui.adsabs.harvard.edu/abs/1994A&A...284..285K} {284, 285}

\bibitem[\protect\citeauthoryear{{Kovetz}}{{Kovetz}}{2017}]{2017PhRvL.119m1301K}
{Kovetz} E.~D.,  2017, \mn@doi [\prl] {10.1103/PhysRevLett.119.131301}, \href
  {https://ui.adsabs.harvard.edu/abs/2017PhRvL.119m1301K} {119, 131301}

\bibitem[\protect\citeauthoryear{{LIGO Scientific Collaboration} et~al.,}{{LIGO
  Scientific Collaboration} et~al.}{2015}]{2015CQGra..32g4001L}
{LIGO Scientific Collaboration} et~al., 2015, \mn@doi [Classical and Quantum
  Gravity] {10.1088/0264-9381/32/7/074001}, \href
  {https://ui.adsabs.harvard.edu/abs/2015CQGra..32g4001L} {32, 074001}

\bibitem[\protect\citeauthoryear{{Lai}, {Hannuksela}, {Herrera-Mart{\'\i}n},
  {Diego}, {Broadhurst}  \& {Li}}{{Lai} et~al.}{2018}]{2018PhRvD..98h3005L}
{Lai} K.-H.,  {Hannuksela} O.~A.,  {Herrera-Mart{\'\i}n} A.,  {Diego} J.~M.,
  {Broadhurst} T.,   {Li} T. G.~F.,  2018, \mn@doi [\prd]
  {10.1103/PhysRevD.98.083005}, \href
  {https://ui.adsabs.harvard.edu/abs/2018PhRvD..98h3005L} {98, 083005}

\bibitem[\protect\citeauthoryear{{Lawrence}}{{Lawrence}}{1971}]{1971NCimB...6..225L}
{Lawrence} J.~K.,  1971, \mn@doi [Nuovo Cimento B Serie] {10.1007/BF02735388},
  \href {https://ui.adsabs.harvard.edu/abs/1971NCimB...6..225L} {6B, 225}

\bibitem[\protect\citeauthoryear{{Li}, {Lo}, {Sachdev}, {Chan}, {Lin}, {Li}  \&
  {Weinstein}}{{Li} et~al.}{2019}]{2019arXiv190406020L}
{Li} A. K.~Y.,  {Lo} R. K.~L.,  {Sachdev} S.,  {Chan} C.~L.,  {Lin} E.~T.,
  {Li} T. G.~F.,   {Weinstein} A.~J.,  2019, arXiv e-prints, \href
  {https://ui.adsabs.harvard.edu/abs/2019arXiv190406020L} {p. arXiv:1904.06020}

\bibitem[\protect\citeauthoryear{{Lo} \& {Magana Hernandez}}{{Lo} \& {Magana
  Hernandez}}{2021}]{2021arXiv210409339L}
{Lo} R. K.~L.,  {Magana Hernandez} I.,  2021, arXiv e-prints, \href
  {https://ui.adsabs.harvard.edu/abs/2021arXiv210409339L} {p. arXiv:2104.09339}

\bibitem[\protect\citeauthoryear{{Maggiore} et~al.,}{{Maggiore}
  et~al.}{2020}]{2020JCAP...03..050M}
{Maggiore} M.,  et~al., 2020, \mn@doi [\jcap] {10.1088/1475-7516/2020/03/050},
  \href {https://ui.adsabs.harvard.edu/abs/2020JCAP...03..050M} {2020, 050}

\bibitem[\protect\citeauthoryear{{McIsaac}, {Keitel}, {Collett}, {Harry},
  {Mozzon}, {Edy}  \& {Bacon}}{{McIsaac} et~al.}{2020}]{2020PhRvD.102h4031M}
{McIsaac} C.,  {Keitel} D.,  {Collett} T.,  {Harry} I.,  {Mozzon} S.,  {Edy}
  O.,   {Bacon} D.,  2020, \mn@doi [\prd] {10.1103/PhysRevD.102.084031}, \href
  {https://ui.adsabs.harvard.edu/abs/2020PhRvD.102h4031M} {102, 084031}

\bibitem[\protect\citeauthoryear{{Meena} \& {Bagla}}{{Meena} \&
  {Bagla}}{2020}]{2020MNRAS.492.1127M}
{Meena} A.~K.,  {Bagla} J.~S.,  2020, \mn@doi [\mnras] {10.1093/mnras/stz3509},
  \href {https://ui.adsabs.harvard.edu/abs/2020MNRAS.492.1127M} {492, 1127}

\bibitem[\protect\citeauthoryear{{Mishra}, {Meena}, {More}, {Bose}  \&
  {Bagla}}{{Mishra} et~al.}{2021}]{2021MNRAS.508.4869M}
{Mishra} A.,  {Meena} A.~K.,  {More} A.,  {Bose} S.,   {Bagla} J.~S.,  2021,
  \mn@doi [\mnras] {10.1093/mnras/stab2875}, \href
  {https://ui.adsabs.harvard.edu/abs/2021MNRAS.508.4869M} {508, 4869}

\bibitem[\protect\citeauthoryear{{More} \& {More}}{{More} \&
  {More}}{2021}]{2021arXiv211103091M}
{More} A.,  {More} S.,  2021, arXiv e-prints, \href
  {https://ui.adsabs.harvard.edu/abs/2021arXiv211103091M} {p. arXiv:2111.03091}

\bibitem[\protect\citeauthoryear{{Nakamura}}{{Nakamura}}{1998}]{1998PhRvL..80.1138N}
{Nakamura} T.~T.,  1998, \mn@doi [\prl] {10.1103/PhysRevLett.80.1138}, \href
  {https://ui.adsabs.harvard.edu/abs/1998PhRvL..80.1138N} {80, 1138}

\bibitem[\protect\citeauthoryear{{Nakamura} \& {Deguchi}}{{Nakamura} \&
  {Deguchi}}{1999}]{1999PThPS.133..137N}
{Nakamura} T.~T.,  {Deguchi} S.,  1999, \mn@doi [Progress of Theoretical
  Physics Supplement] {10.1143/PTPS.133.137}, \href
  {https://ui.adsabs.harvard.edu/abs/1999PThPS.133..137N} {133, 137}

\bibitem[\protect\citeauthoryear{{Narayan} \& {Bartelmann}}{{Narayan} \&
  {Bartelmann}}{1996}]{1996astro.ph..6001N}
{Narayan} R.,  {Bartelmann} M.,  1996, arXiv e-prints, \href
  {https://ui.adsabs.harvard.edu/abs/1996astro.ph..6001N} {pp
  astro--ph/9606001}

\bibitem[\protect\citeauthoryear{{Oguri}}{{Oguri}}{2018}]{2018MNRAS.480.3842O}
{Oguri} M.,  2018, \mn@doi [\mnras] {10.1093/mnras/sty2145}, \href
  {https://ui.adsabs.harvard.edu/abs/2018MNRAS.480.3842O} {480, 3842}

\bibitem[\protect\citeauthoryear{{Ohanian}}{{Ohanian}}{1974}]{1974IJTP....9..425O}
{Ohanian} H.~C.,  1974, \mn@doi [International Journal of Theoretical Physics]
  {10.1007/BF01810927}, \href
  {https://ui.adsabs.harvard.edu/abs/1974IJTP....9..425O} {9, 425}

\bibitem[\protect\citeauthoryear{{Padilla} \& {Strauss}}{{Padilla} \&
  {Strauss}}{2008}]{2008MNRAS.388.1321P}
{Padilla} N.~D.,  {Strauss} M.~A.,  2008, \mn@doi [\mnras]
  {10.1111/j.1365-2966.2008.13480.x}, \href
  {https://ui.adsabs.harvard.edu/abs/2008MNRAS.388.1321P} {388, 1321}

\bibitem[\protect\citeauthoryear{{Saha} \& {Williams}}{{Saha} \&
  {Williams}}{2011}]{2011MNRAS.411.1671S}
{Saha} P.,  {Williams} L. L.~R.,  2011, \mn@doi [\mnras]
  {10.1111/j.1365-2966.2010.17797.x}, \href
  {https://ui.adsabs.harvard.edu/abs/2011MNRAS.411.1671S} {411, 1671}

\bibitem[\protect\citeauthoryear{{Salpeter}}{{Salpeter}}{1955}]{1955ApJ...121..161S}
{Salpeter} E.~E.,  1955, \mn@doi [\apj] {10.1086/145971}, \href
  {https://ui.adsabs.harvard.edu/abs/1955ApJ...121..161S} {121, 161}

\bibitem[\protect\citeauthoryear{{Schneider}, {Ehlers}  \& {Falco}}{{Schneider}
  et~al.}{1992}]{1992grle.book.....S}
{Schneider} P.,  {Ehlers} J.,   {Falco} E.~E.,  1992, {Gravitational Lenses},
  \mn@doi{10.1007/978-3-662-03758-4.
}

\bibitem[\protect\citeauthoryear{{Seo}, {Hannuksela}  \& {Li}}{{Seo}
  et~al.}{2021}]{2021arXiv211003308S}
{Seo} E.,  {Hannuksela} O.~A.,   {Li} T. G.~F.,  2021, arXiv e-prints, \href
  {https://ui.adsabs.harvard.edu/abs/2021arXiv211003308S} {p. arXiv:2110.03308}

\bibitem[\protect\citeauthoryear{{Takahashi} \& {Nakamura}}{{Takahashi} \&
  {Nakamura}}{2003}]{2003ApJ...595.1039T}
{Takahashi} R.,  {Nakamura} T.,  2003, \mn@doi [\apj] {10.1086/377430}, \href
  {https://ui.adsabs.harvard.edu/abs/2003ApJ...595.1039T} {595, 1039}

\bibitem[\protect\citeauthoryear{{The LIGO Scientific Collaboration}
  et~al.,}{{The LIGO Scientific Collaboration}
  et~al.}{2021}]{2021arXiv211103606T}
{The LIGO Scientific Collaboration} et~al., 2021, arXiv e-prints, \href
  {https://ui.adsabs.harvard.edu/abs/2021arXiv211103606T} {p. arXiv:2111.03606}

\bibitem[\protect\citeauthoryear{{Ulmer} \& {Goodman}}{{Ulmer} \&
  {Goodman}}{1995}]{1995ApJ...442...67U}
{Ulmer} A.,  {Goodman} J.,  1995, \mn@doi [\apj] {10.1086/175422}, \href
  {https://ui.adsabs.harvard.edu/abs/1995ApJ...442...67U} {442, 67}

\bibitem[\protect\citeauthoryear{{Usman} et~al.,}{{Usman}
  et~al.}{2016}]{2016CQGra..33u5004U}
{Usman} S.~A.,  et~al., 2016, \mn@doi [Classical and Quantum Gravity]
  {10.1088/0264-9381/33/21/215004}, \href
  {https://ui.adsabs.harvard.edu/abs/2016CQGra..33u5004U} {33, 215004}

\bibitem[\protect\citeauthoryear{{Vernardos}}{{Vernardos}}{2019}]{2019MNRAS.483.5583V}
{Vernardos} G.,  2019, \mn@doi [\mnras] {10.1093/mnras/sty3486}, \href
  {https://ui.adsabs.harvard.edu/abs/2019MNRAS.483.5583V} {483, 5583}

\bibitem[\protect\citeauthoryear{{Vijaykumar}, {Mehta}  \&
  {Ganguly}}{{Vijaykumar} et~al.}{2022}]{Vijaykumar2022}
{Vijaykumar} A.,  {Mehta} A.~K.,   {Ganguly} A.,  2022, arXiv e-prints, \href
  {https://ui.adsabs.harvard.edu/abs/2022arXiv220206334V} {p. arXiv:2202.06334}

\makeatother
\end{thebibliography}

%%%%%%%%%%%%%%%%%%%%%%%%%%%%%%%%%%%%%%%%%%%%%%%%%%%%%%%%%%%%
%\appendix

%%%%%%%%%%%%%%%%%%%%%%%%%%%%%%%%%%%%%%%%%%%%%%%%%%%%%%%%%%%%

% Don't change these lines
\bsp	% typesetting comment
\label{lastpage}
\end{document}